\documentclass{aastex62}

\received{\today}
\submitjournal{ApJ}

\shorttitle{Black Hole Spin and Disk Magnetic Field Strength Estimates}
\shortauthors{Daly}
\usepackage{longtable}
\begin{document}

\title{Black Hole Spin and Accretion Disk Magnetic Field Strength Estimates for 
more than 750 AGN and Multiple GBH}

\correspondingauthor{Ruth Daly}
\email{rdaly@psu.edu}

\author[0000-0000-0000-0000]{Ruth A. Daly}

\affil{Penn State University, Berks Campus, Reading, PA 19608, USA}
\affil{Center for Computational Astrophysics, Flatiron Institute, 162 5th Avenue, New York, NY 10010, USA}

\nocollaboration



\begin{abstract}
Black hole systems, comprised of a black hole, accretion disk, and 
collimated outflow are studied here. Three AGN samples including 753 AGN, and 
102 measurements of 4 GBH are studied. Applying the 
theoretical considerations described by Daly (2016), 
general expressions for the black hole spin function 
and accretion disk magnetic field strength are presented and 
applied to obtain the 
black hole spin function, spin, and accretion disk magnetic
field strength in dimensionless and physical units 
for each source.  Relatively high
spin values are obtained; spin functions indicate typical spin values of  
about (0.6 - 1) for the sources. The distribution of accretion 
disk magnetic field strengths for the three AGN 
samples are quite broad and have mean values of about $10^4$ G, 
while those for individual GBH have mean values of about  $10^8$ G. Good 
agreement is found between spin values obtained here and published values 
obtained with well-established methods; comparisons for 1 GBH and 6 AGN 
indicate that similar spin values are obtained with independent methods.
Black hole spin and disk magnetic field strength demographics 
are obtained and indicate that black hole spin  
functions and spins are similar for all of the source types studied 
including GBH and different categories of AGN.
The method applied here does not depend upon any specific accretion 
disk emission model, and does not depend upon a specific model that relates 
jet beam power to compact radio luminosity, hence the results obtained
here can be used to constrain and study these models.
\end{abstract}



\section{Introduction} \label{sec:intro}

Several methods have been proposed to measure and study the spin properties 
of supermassive black holes associated with active galactic nuclei (AGN) 
and stellar mass black holes that are X-ray binaries, 
referred to as Galactic Black Holes (GBH). 
These methods include the use of emission from the accretion disk 
associated with the black hole (Fabian et al. 1989;  Miller et al. 2009;
Patrick et al. 2012; Walton et al. 2013; Vasudevan et al. 2016; 
Reynolds 2019 and references therein); 
the use of the properties of extended radio sources (Daly 2009a,b, 2011,   
Daly \& Sprinkle 2014); 
a combination of accretion disk emission and the properties of
radio sources (Gnedin et al. 2012; Daly 2016; Mikhailov \& Gnedin 2018); 
and the properties of gravitational waves (Abbott et al. 2018). 
To date, studies of emission from the accretion disk alone has led to 
about 22 AGN with ``robust'' spin determination based on the X-ray reflection
method (e.g. Reynolds 2019 and references therein),
and disk emission has led to several GBH spin determinations 
(e.g. Miller et al. 2009; King et al. 2013 and references therein). 
Studies of outflows from powerful extended radio sources has led to about
130 spin determinations (e.g. Daly \& Sprinkle 2014), 
while studies that combine the disk emission and outflow properties
provide a few hundred spin values (e.g. Gnedin et al. 2012; 
Daly 2016; Mikhailov \& Gnedin 2018). 

Here, accretion disk emission properties are combined with the 
properties of radio sources associated with the collimated outflow from 
a black hole to estimate black hole spin functions
and spins for 753 active galactic nuclei (AGN) and 4 Galactic Black 
Holes (GBH) using the method proposed by Daly (2016) and extended here.
The ``black hole system'' includes 
the black hole, the accretion disk, and the collimated outflow.  
Background information on measuring black hole spin for sources
with collimated outflows is provided in section 1.1.

In addition to providing a black hole spin estimate, the 
method introduced by Daly (2016) and extended here 
allows an estimate of the magnetic field strength of
the accretion disk of each source in a manner that does not depend
upon a detailed model of the accretion 
disk. Here and throughout the paper, the term ``magnetic field strength'' 
indicates the magnitude of the magnetic field, $B$, of the accretion disk. 

\subsection{Background on Black Hole Systems with Collimated Outflows}

To empirically determine and study black hole spin, sources for which the
spin is likely to affect some empirically accessible 
aspect of the source are selected for study.
Black hole systems with powerful outflows, especially those with
outflows that are powerful relative to the 
accretion disk bolometric luminosity, are likely to be powered at least in 
part by the spin of the black hole (e.g. Blandford \& Znajek 1977; 
Begelman, Blandford, \& Rees 1984; Blandford 1990; Meier 1999, 2001; 
Blandford, Meier, \& Readhead 2018; Yuan \& Narayan 2014). This is based 
on the premise that the spin energy of the black hole
can be extracted (Penrose 1969; Penrose \& Floyd 1971). 
 
Thus, sources with powerful collimated 
outflows are considered here. The beam power, $L_j \equiv dE/dt$, 
is the energy per unit time ejected by the black hole system 
in the form of a collimated jet. The beam power is typically considered to be 
in the form of directed kinetic energy 
and the jet manifests its presence as a compact or extended radio source 
The beam powers of classical double radio sources discussed here 
were obtained with the ``strong shock method,'' described in detail below. 

A source with both a beam power and black 
hole mass determination can be used to solve for a combination of
the black hole spin and accretion disk braking magnetic field
strength (Daly 2009a,b, 2011; Daly \& Sprinkle 2014; Mikhailov \& Gnedin 2018). 
For example, Daly (2011) and Daly \& Sprinkle (2014) 
assumed three different parameterizations of the magnetic field strength
and solved for black hole spin values given 
these field strengths. Thus, a second equation is needed to be
able to break this magnetic field strength - black hole spin  
degeneracy and solve for the spin and accretion disk 
magnetic field strength separately. 

To be able to break this degeneracy, 
Daly (2016) (hereafter D16) considered sources that have 
bolometric accretion disk
luminosity determinations in addition to values for the beam power 
and black hole mass. This led to a sample of 
97 powerful classical double radio sources.  
This study indicated an empirically determined 
relationship between the beam power, accretion disk bolometric luminosity, 
$L_{bol}$, and Eddington luminosity, $L_{Edd}$, of the form 
\begin{equation}
{L_j \over L_{bol}} \propto \left({L_{bol} \over L_{Edd}}\right)^{\alpha_*} 
 \propto \left({L_{bol} \over L_{Edd}}\right)^{A-1}, 
\end{equation}
where the parameter $\alpha_*$ is related to the parameter $A$ by 
$\alpha_* = (A-1)$; a best fit value of 
$\alpha_* = -0.56 \pm 0.05$ was obtained. Eq. (1) is referred to as the 
fundamental line (FL) of black hole activity. The Eddington luminosity 
is given by $L_{Edd} \simeq 1.3 \times 10^{46} M_8$ erg/s, where $M_8$ is the 
black hole mass in units of $10^8 M_{\odot}$.  
The 97 sources included in the study are powerful classical double
radio sources, known as FRII sources (Fanaroff \& Riley 1974). 

The beam powers were determined using the ``strong shock'' method; 
the method is based on the application of the equations of strong shock physics 
(i.e. the strong shock jump conditions) to powerful classical double sources 
(Daly 1994). The method can only be applied to FRII radio sources 
with very regular radio bridge or lobe structure, which corresponds to 
sources with 178 MHz radio powers greater than about $8 \times 10^{27}$ W/Hz 
for a value of Hubble's constant of 70 km/s/Mpc 
(Leahy \& Williams 1984; Leahy, Muxlow, \& Stephens 1989; Wellman, Daly, 
\& Wan 1997a,b; O'Dea et al. 2009). The cigar-like lobe structure indicates
that the sources are not self-similar, which is also indicated by 
detailed studies of the source shapes (e.g. Leahy \& Williams 1984; 
Leahy, Muxlow, \& Stephens 1989; Wellman et al. 1997a; Daly et al. 2010). 
The method requires the use of high resolution 
multifrequency radio maps covering the full radio lobe region of each source. 
O'Dea et al. (2009) showed that this method is essentially model
independent: it has no free parameters and 
does {\it not} rely on whether the radio emitting
plasma is close to minimum energy or equipartition conditions 
when the magnetic field strength in the extended radio lobe is
less than or equal to the minimum energy value (see section 3.3 of O'Dea et
al. 2009), which is consistent with measured offsets from 
minimum energy conditions in sources of this type 
(e.g. Wellman et al. 1997a,b; Croston et al. 2004, 2005; 
Shelton, Hardcastle, \& Croston 2011; Godfrey \& Shabala 2013; 
Harwood et al. 2016).  When the beam power is 
written in terms of empirically determined strong shock parameters, 
the parameterized deviations of the relativistic electron population and
magnetic field strength of the plasma in the extended radio source 
from minimum energy (or equipartition) conditions fortuitously cancel out.  
Beam powers for additional sources 
were obtained by applying the relationship between beam power obtained with 
the strong shock method and radio power (Daly 
et al. 2012). The beam powers obtained for the FRII sources 
discussed here using the strong shock method are thought to be very reliable. 
From a strictly empirical perspective, these beam powers are the 
foundation of the use of FRII sources for cosmological studies, and
are found to yield coordinate distances, and first and second derivatives 
of the coordinate distance with respect to redshift that are in excellent 
agreement with those obtained using type Ia supernovae 
from a redshift of about zero to a redshift 
greater than one  (e.g. Daly et al. 2008).
The black hole masses were obtained from McLure et al. (2006), 
McLure et al. (2004), and Tadhunter et al. (2003). 
The bolometric accretion disk luminosities,
$L_{bol}$ were obtained from the [OIII] luminosities using the 
relationship obtained by Heckman et al. (2004) and confirmed
by Dicken et al. (2014). The [OIII] luminosities were obtained from
Grimes, Rawlings, \& Willott (2004). 

To understand the implications of the 
empirical relationship given by eq. (1), general theoretical 
expressions were considered.
The beam power is parameterized as $L_j \propto \dot{m}^a ~M^b~f(j)$, 
where $M$ is the black hole mass,
$f(j)$ is a function of the spin of the black hole, and the 
bolometric disk luminosity is parameterized as
$L_{bol} \propto \epsilon \dot{M} \propto \epsilon ~ \dot{m} ~M$, where
$\dot{M}$ is the mass accretion rate, $\dot{m} \equiv \dot{M}/\dot{M}_{Edd}$
is the dimensionless mass accretion rate, 
$\dot{M}_{Edd} \equiv L_{Edd} c^{-2}$ is the Eddington accretion rate,
and $\epsilon$ is a dimensionless efficiency factor. The 
efficiency $\epsilon$ and the dimensionless mass accretion rate $\dot{m}$ 
are each normalized to have a maximum value of unity.
The spin function $f(j)$ is considered to be independent
of the mass accretion rate and black hole mass. 
Here, the notation of Blandford (1990) and Meier (1999) is used;
the dimensionless black hole spin $j$ is defined in the usual way, 
$j \equiv Jc/(GM^2)$, where $J$ is the spin angular momentum of the hole, 
$M$ is the black hole mass, $c$ is the speed of light, and $G$ is 
Newton's constant; in other work, $j$ is sometimes represented with the 
symbol $a$ or $a_*$. The Meier (1999) model is a hybrid that 
includes a disk wind, as in 
the model of Blandford \& Payne (1982), and black hole 
spin energy extraction, as in the model of Blandford \& Znajek (1977).
The normalizations are parameterized by $g_j$ and $g_{bol}$ where
the maximum values of the beam 
power and bolometric luminosities are given by $L_j(max) = g_j L_{Edd}$
and $L_{bol}(max) = g_{bol} L_{Edd}$, respectively. 
This form for $L_j$ is motivated by the form of the equation for the beam
power in black hole spin powered outflow models: $L_j \propto \dot{m} ~M~f(j)$ 
$\propto B_p^2 ~M^2~f(j)$, that is, $B^2_p \propto  (\dot{m}/M)$ 
(e.g. Blandford \& Znajek 1977; Blandford 1990; Meier 1999; 
Tchekhovskoy, Narayan, \& McKinney 2010; Yuan \& Narayan 2014), 
where $B_p$ is the poloidal component of the 
accretion disk magnetic field. A spin term is not included in the expression for
$L_{bol}$ because this is expected to be important only as 
$L_{bol}/L_{Edd} \rightarrow 1$, which would only impact a small fraction
of the sources considered;  
as the sample sizes become larger, it will be possible 
to look for the signature of the impact of spin on $L_{bol}$. 

The theoretical expressions are applied 
to the  empirical relationships to understand the implications of the results.  
Eq. (1) indicates that $\dot{m}^a ~M^b~f(j) \propto (\epsilon ~\dot{m})^{A} M$, 
which suggests that $b=1$. In this case, the ratio of $L_j/L_{bol}$ is 
expected to be independent of black hole mass, 
which is confirmed to be the case for the sample studied by D16. 
A value of $b=1$, indicates that 
$\dot{m}^a ~M~f(j) \propto (\epsilon ~\dot{m})^{A} M$, so  
$\dot{m}^a = (\epsilon ~\dot{m})^{A}$, since $f(j)$ 
is independent of $\dot{m}$ and $M$ by construction, and
$\epsilon$ and $\dot{m}$ are each normalized to have a maximum value of one. 
While this general solution leads to eq. (2), 
it is interesting to pause for a moment and 
consider the two simplest particular solutions for 
$A \simeq 1/2$ (obtained by D16): that  
obtained with $a=1$, indicating $\epsilon \propto \dot{m}$, 
and that obtained for $a=1/2$, indicating $\epsilon$ = constant. 
The first particular solution indicates $L_j \propto \dot{m} ~M~f(j)$
with $L_{bol} \propto \dot{m}^2 ~M$. The second particular solution indicates
$L_j \propto \dot{m}^{1/2} ~M~f(j)$ with $L_{bol} \propto \dot{m} ~M$. 
This means that the combination $L_j \propto \dot{m} ~M~f(j)$ 
and $L_{bol} \propto \dot{m} ~M$ is inconsistent with the empirical 
results obtained by D16.  

The general solution obtained above indicates that 
(see eqs. 4 and 5 of D16),  
independent of the value of $a$, 
\begin{equation}
{f(j) \over f_{max}} = \left({L_{j} \over g_j L_{Edd}}\right)~\left({
L_{bol} \over g_{bol} L_{Edd}}\right)^{-A}~.
\end{equation}
The spin function $f(j)$ is normalized by 
its maximum value $f_{max}$, which is the value of $f(j)$ when $j = 1$. 

It will be shown in section 3 that the same expression
can be derived for other sources that fall on the fundamental line of 
black hole activity. The spin function
can be converted to black hole spin $j$, as discussed in section 4.
The quantity $\sqrt{(f(j)/f_{max})} \propto j$ to first
order in most outflow models, 
so this is the quantity obtained and studied here. In the Blandford-Znajek
model (1977) and the Meier (1999) models, the conversion is 
$\sqrt{f(j)/f_{max}} = j (1+\sqrt{1-j^2})^{-1}$, though numerical simulations
suggest that this equation may be modified for some outflow models
(e.g. Tchekhovskoy et al. 2010, Yuan \& Narayan 2014). 

Given the results reported by D16, it was interesting to consider whether
the fundamental plane of black hole activity, which is written in terms of 
observed quantities, is the empirical manifestation of the same 
underlying relationship as that described by eq. (1). Daly et al. (2018) 
(hereafter D18) showed that this is indeed the case.

There are many equivalent representations of the fundamental plane (FP), 
which is a relationship that applies to sources with 
{\it compact} radio emission, disk X-ray emission, and an estimate of the 
black hole mass. 
The FP was introduced by Merloni, Heinz, \& De Matteo (2003) (hereafter M03) 
and Falcke, K\"ording, \& Markoff (2004).  
Here, the representation presented by Nisbet \& Best (2016) (hereafter NB16)
is followed; 
other representations can easily be obtained from this by making
appropriate substitutions as described by NB16:  
\begin{equation}
\rm{log} ~L_R = a~\rm{log}~L_{X,42}  + b~\rm{log}~ M_8 + c~,
\end{equation}
where $a$, $b$, and $c$ are empirically determined constants 
and may be different for different source samples
(note, these are unrelated to the quantities $a$ and $b$ discussed 
earlier); 
$L_R$ is the 1.4 GHz radio power of the compact radio source in 
$\hbox{erg s}^{-1}$; $L_{X,42}$ is the (2 - 10) keV X-ray luminosity of 
the disk in units of 
$10^{42} \hbox{ erg s}^{-1}$; and $M_8$ is the black hole mass in units
of $10^8 M_{\odot}$; conversions to and from other wavebands are discussed by 
NB16. NB16 present results obtained with a sample 
of 576 LINERS and summarize results obtained by M03, 
K\"ording et al. (2006a,b), G\"ultekin et al. (2009), Bonchi 
et al. (2013), and Saikia et al. (2015) (hereafter S15). 

For FP sources, 
D18 showed that using standard, well-accepted equations 
relating the compact radio luminosity to the beam power, and the
disk X-ray luminosity to the bolometric disk luminosity, the FP can
be written in the form of eq. (1) without specifying a value of 
$\alpha_*$ or $A$, and best fit FP parameters
can be used to obtain the beam power from the compact radio luminosity.
Specifically, the beam power is written as 
\begin{equation}
\rm{log} L_j = C ~\rm{log} ~L_R  + D ~,
\end{equation}
and the bolometric disk luminosity as
\begin{equation}
L_{bol} = \kappa_{X} ~ L_{X} (2-10 \hbox{ keV}),
\end{equation}
where $C$, $D$, and $\kappa_X$ are constant for a given source sample, 
and $L_{X} (2-10 \hbox{ keV})$ is the (2-10) keV luminosity of the source. 
Given eqs. (4) and (5), D18 showed that eq. (3) can be written in the form of 
eq. (1). The results indicate that the value of $C$ is
related to the best fit FP parameters $a$ and $b$: $C = 1/(a+b)$; 
$D$ depends upon a series of empirically determined 
constants, including best fit FP parameters $a$, $b$, and $c$,  
$\kappa_X$, and a normalization factor, $B$, obtained using the strong
shock method, as described in detail by D18. 

Thus, the beam power $L_j$ can be obtained empirically in a completely 
model-independent manner for sources that lie on the FP, using 
eq. (4) once the values of $C$ and $D$ have been obtained from 
best fit FP values $a$, $b$, and $c$ for that sample of sources. 
The beam powers for the FP samples considered here are obtained
using the values of $C$ and $D$ listed in columns (2) and (9) of Table 1 
of D18. No specific detailed 
model relating the compact radio luminosity to beam power is required to obtain
these beam powers. The normalization from the FRII beam powers does enter
through the parameter $B$ determined with FRII AGN (see D16 and D18). 
Because the normalization of the beam powers for FP sources is tied to 
that for the FRII sources, any shift caused by an offset of this normalization
from the true value affects all of the sources in the same manner. This will
be important later when the overall normalization of the beam powers, $g_j$,
is discussed. Any normalization offset of the empirically determined 
beam powers from the true value will be absorbed by the normalization 
factor $g_j$, since it is the quantity $(L_j/g_j)$ that enters into the 
determination of the black hole spin function.  

Several authors have investigated whether the X-ray emission
associated with FP sources is disk or jet emission (e.g. see the 
summary by Yuan \& Narayan 2014). Both empirical
and theoretical studies indicate that the X-ray emission arises
from the accretion disk emission except in cases 
of very low X-ray luminosities, luminosities in Eddington units of
less than about $10^{-6}$ (M03; Yuan \& Cui 2005; Yuan \& Narayan 2014; 
Xie \& Yuan 2017); for the sources studied here, this could affect the
results for only a few sources, notably for Sgr A$^*$. 
In addition, S15 studied FP sources using [OIII] 
luminosity rather than X-ray luminosity, and found the same best fit 
FP parameters, indicating that the X-ray emission  
originates from the accretion disk. Finally, the slope of the FL is very 
similar for FP sources (D18) and FRII AGN (D16), and the 
bolometric luminosities of the FRII AGN are based on [OIII] 
luminosities. The value of the bolometric luminosity of the disk is obtained
using eq. (5) with $\kappa_X = 15.1$  (Ho 2009; D18). 
Given that all of the sources on the FP have black hole mass
determinations, it is easy to construct the FL (given by eq. 1) 
for separate samples, as was done by D18 for the NB16, S15, and M03
FP sources plus the FL sources studied by D16. 

Here, the methods introduced by D16 are applied and extended to the four 
data sets for which the FL of black hole activity was studied 
(D16; D18). The application of these methods allows  
estimates of the spin function and spin of each black hole, and the
accretion disk magnetic field strength in dimensionless and physical
units of each source. 
Since these rely upon the beam power, bolometric accretion disk luminosity,
and black hole mass (or Eddington luminosity), these quantities are 
also included in the tables that list values of all quantities, 
and are discussed in section 2. 
Given the large numbers of sources included in the study, black hole
demographics can be studied.    

The data used for these studies is discussed in section 2. 
The methods of obtaining the spin function, black hole spin, 
and accretion disk magnetic field strength are 
discussed in section 3. The results are presented in section 4, 
discussed in section 5, and summarized in section 6. 

\section{The Data}

All quantities are obtained in a spatially flat cosmological model with 
two components, a mean mass density relative to the critical value at the 
current epoch of $\Omega_m = 0.3$ and a cosmological constant normalized
by the critical density today of $\Omega_{\Lambda} = 0.7$; a value
for Hubble's constant of $H_0 = 70$ km/s/Mpc is assumed throughout.

The data sets studied here include the 576 LINERS from NB16, 
the 102 data points for 4 Galactic Black Holes (GBH) listed by 
S15, the 97 FRII sources from D16, and the 80 AGN from M03.   
All four samples were studied in detail by D18. With the exception
of a few black hole mass updates, the radio powers, 
X-ray luminosities, and black hole masses in the 
published source tables of NB16, S15, and M03 are used to obtain beam
powers, bolometric luminosities, and Eddington luminosities as described in 
section 1.1. For the sample of D16, these quantities are obtained as
described in section 1.1. 
Black hole masses have been updated for Sgr $A^*$ (Ghez et al. 2008), 
Ark 564 (Vasudevan et al. 2016; Ilic et al. 2017), 
Mrk 335 (Grier et al. 2017), 
NGC 4051 (Seifina et al. 2018), NGC 4151 (Bentz et al. 2006), 
3C 120 (Grier et al. 2017), NGC 1365 (Fazeli et al. 2019), 
GX 339-4 (King et al. 2013), XTEJ118+480 (Khargharia et al. 2013), 
and AO6200 (Cantrell et al. 2010). 

Note that the beam powers are obtained using the strong
shock method for the D16 sources, while they are obtained using the 
``CD'' method (given by eq. 4) for the three other samples.  
The bolometric disk luminosities are obtained from the 
[OIII] luminosity for the D16 sources while they are obtained from 
X-ray luminosities for the three other samples. 
The sources studied by D16 are powerful extended classical
double sources, while the radio sources in the other three samples 
are selected to have a powerful compact radio component. 
Yet all of the sources lie on the FL and have very 
similar best fit FL parameters (D18). 
Interestingly, S15 showed that equivalent FP results
are obtained when [OIII] luminosity rather than X-ray luminosity 
is used as an indicator of the bolometric luminosity of AGN. 

A significant number of sources studied by D18 
have beam powers one to two orders of magnitude
larger than the bolometric luminosity of the disk. 
For these sources, 
it is reasonable to suppose that the outflow is powered, at least in part,
by the spin of the black hole. Since the slope of the relationship
indicated by eq. (1) 
is nearly identical for all samples studied, it is
reasonable to conclude that all of the sources are governed by the same 
physics, and, hence that the outflow of each of the sources is  
powered, at least in part, by the spin of the hole. 

\begin{figure}
    \centering
    \includegraphics[width=80mm]{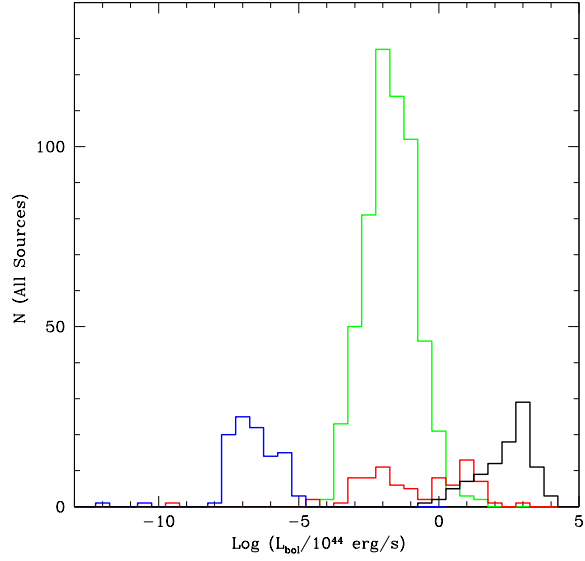}
\caption{Histograms of the bolometric luminosity are
shown for each of the four samples studied here. Here and throughout the 
paper, the 576 LINERS from NB16 are shown in green; the 97 FRII sources 
from D16 are shown in black; the 80 AGN from M03 are shown in red; 
and the 102 GBH from S15 are shown in blue. 
The mean value and standard deviations of the distributions are: 
$-1.69 \pm 0.90$ (576 LINERS from NB16); 
$2.36 \pm 0.97$ (97 FRII sources from D16); 
$-0.89 \pm 1.99$ (80 compact radio sources from M03); and 
$-6.64 \pm 0.97$ (102 GBH from S15).}
		  \label{fig:newF1}
    \end{figure}

\begin{figure}
    \centering
    \includegraphics[width=80mm]{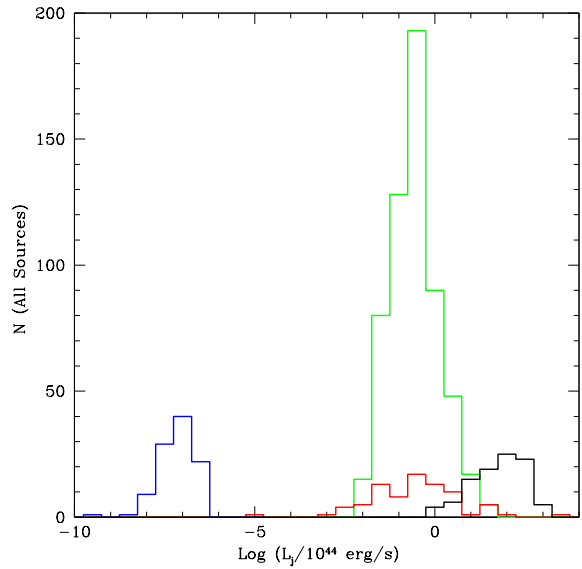}
\caption{Histograms of the beam power are
shown for each of the four samples studied here; the sample colors are 
as in Fig. 1. The mean value and standard deviations of the distributions are: 
$-0.55 \pm 0.67$ (576 LINERS from NB16); 
$1.71 \pm 0.75$ (97 FRII sources from D16); 
$-0.54 \pm 1.25$ (80 compact radio sources from M03); and 
$-7.15 \pm 0.49$ (102 GBH from S15).}
		  \label{fig:newF2}
    \end{figure}

\begin{figure}
    \centering
    \includegraphics[width=80mm]{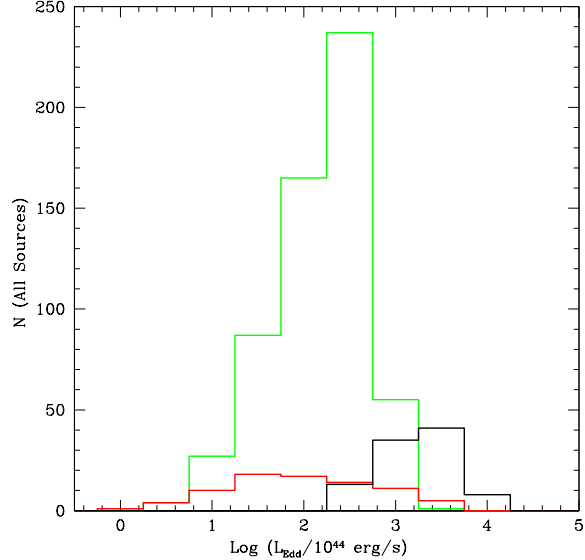}
\caption{Histograms of the Eddington luminosity are shown for each of 
the three AGN samples studied here; the sample colors are as in Fig. 1. 
The mean value and standard deviations of the distributions are: 
$2.19 \pm 0.50$ (576 LINERS from NB16); 
$3.24 \pm 0.40$ (97 FRII sources from D16); 
$1.95 \pm 0.81$ (80 compact radio sources from M03); and 
$-5.07 \pm 0.09$ (102 GBH from S15), though these are not
shown in the figure.}
		  \label{fig:newF4}
    \end{figure} 

\begin{figure}
    \centering
    \includegraphics[width=80mm]{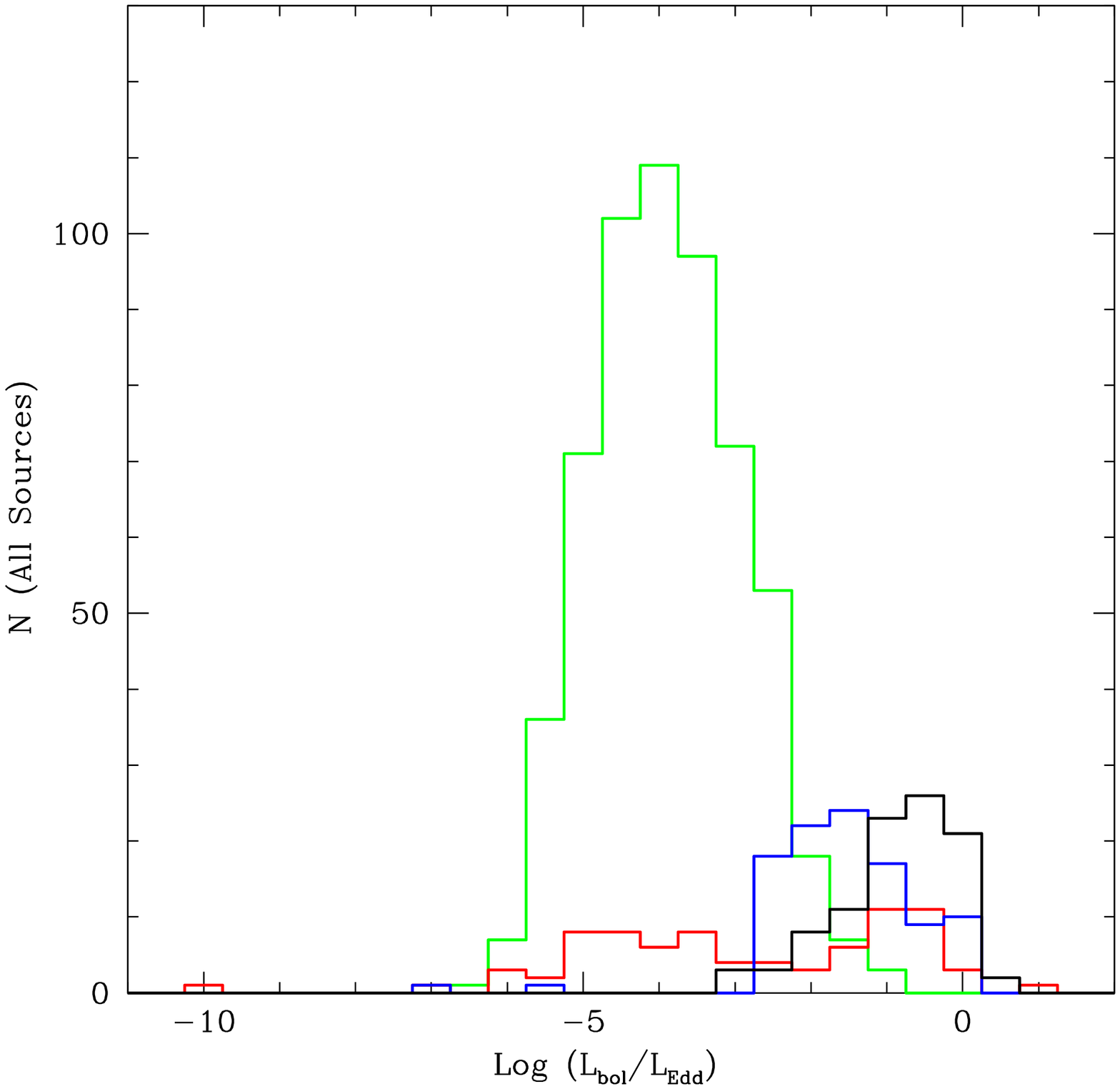}
\caption{Histograms of the Eddington-normalized bolometric luminosity are
shown for each of the four samples studied here; the sample colors are 
as in Fig. 1. The mean value and standard deviations of the distributions are: 
$-3.88 \pm 0.98$ (576 LINERS from NB16); 
$-0.88 \pm 0.77$ (97 FRII sources from D16); 
$-2.84 \pm 2.06$ (80 compact radio sources from M03); and 
$-1.57 \pm 0.98$ (102 GBH from S15).}
		  \label{fig:newF1}
    \end{figure} 

The key quantities used to solve for parameter values characterizing the disk
magnetic field strength and black hole spin of each source are the 
bolometric disk luminosity, $L_{bol}$, beam power, $L_j$, and black hole mass, 
parameterized by the Eddington luminosity, $L_{Edd}$. 
All quantities are obtained as described in section 1.1.
Histograms of $L_{bol}$, $L_j$, $L_{Edd}$, $L_{bol}/L_{Edd}$, $L_j/L_{Edd}$, 
$L_{Tot}/L_{Edd}$, and $L_j/L_{bol}$, are shown in 
Figs. 1 - 7; $L_{Tot} = L_{bol} + L_j$ represents the total power produced 
by the source. 

The uncertainties of $L_{bol}$, $L_j$, and $L_{Edd}$ factor into
the uncertainties of the quantities discussed in sections 1.1, 
3, 4, and 5. The uncertainty of the log of the 
bolometric accretion disk luminosity, $\delta \rm{Log}(L_{bol})$,
obtained from the [OIII] luminosity, applicable to the FRII sources, 
is about 0.38 (Heckman et al. 2004). And, $\delta \rm{Log}(L_{bol})$ for 
FP sources obtained from the (2-10) keV luminosity is about $0.27$ (D18), which 
is obtained by multiplying the uncertainty of the mean of $0.03$ by the 
square root of the number of sources (80) included in the study to obtain
the uncertainty per source. The uncertainties of the beam 
powers for the representative sample of FRII sources listed 
in O'Dea et al. (2009) are approximately $(\delta L_j)/L_j \simeq 0.3$, 
which translates to an uncertainty of the log of the beam power, 
$\delta \rm{Log}(L_j) \simeq 0.13$. Values of  $\delta \rm{Log}(L_j)$ 
for the FP sources obtained as described by D18 are estimated to be about 
0.16 for the LINERS, 0.24 for the M03 sources, and 0.07 
for the GBH; these were estimated by scaling the uncertainty of 0.13 for  
FRII AGN by the ratio of the FL dispersions listed in column (5) of Table 2 
(D18) relative to the value for FRII AGN. Black hole masses are 
obtained using different methods for each of the four samples studied.  
For the NB16 LINERS, the black hole masses are obtained using the SDSS galaxy velocity
dispersion $\sigma$ and applying the McConnell \& Ma (2013) relation to obtain
the black hole mass of each source; the  uncertainty is estimated to
be about $\delta \rm{Log}(L_{Edd}) \simeq 0.3$ to 0.4 for typical
masses of these black holes, with the uncertainty increasing with black hole mass. 
For the M03 sample, the black hole masses are obtained 
using the relationship between black hole mass and galaxy velocity 
dispersion from Ferrarese \& Merritt (2000) and updated by 
Ferrarese (2002), which has an uncertainty of about $\delta \rm{Log}(L_{Edd}) 
\simeq 0.5$. For the FRII radio sources, $\delta \rm{Log}(L_{Edd})$ is about:  
$0.3$ for the radio galaxies obtained from McLure et al. (2004); about $0.3$ for a source
at a redshift of about 1 for the radio galaxies obtained from 
McLure et al. (2006); and $0.4$ for the radio loud quasars 
obtained from McLure et al. (2006). Note that for any quantity x, 
$\delta \rm{Log}(x) = (\delta x/x)(1/\rm{Ln}(10))$, so adding fractional
uncertainties such as $(\delta x/x)$ in quadrature to obtain a total uncertainty
translates into adding terms such as $\delta \rm{Log}(x)$ in quadrature to 
obtain the total uncertainty of the log of a quantity that depends upon 
more than one factor. For the GBH a mass uncertainty 
of about $\delta \rm{Log}(L_{Edd}) \simeq 0.037$ is indicated by 
Khargharia et al. (2013) for XTEJ118+480, and 
$\delta \rm{Log}(L_{Edd}) \simeq 0.016$ is indicated by 
Cantrell et al. (2010) for AO6200; in addition, when considering data
on one particular GBH, the variation of the mass term should be set to zero.
These uncertainties of the mass are quite small relative to the 
uncertainties of the beam power and disk bolometric luminosity, and 
thus can be ignored for the GBH.

\begin{figure}
    \centering
    \includegraphics[width=80mm]{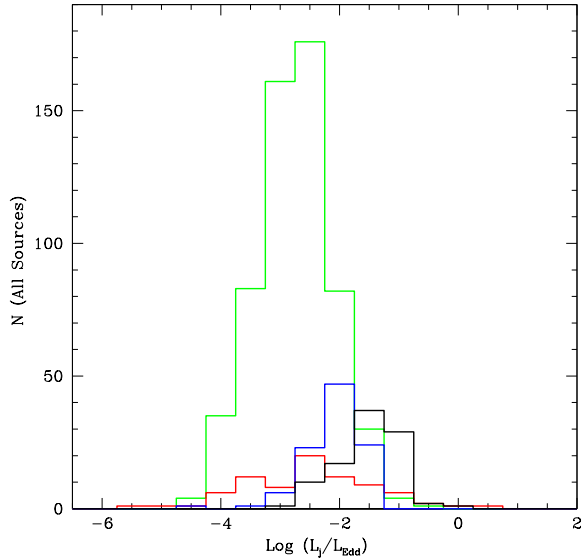}
\caption{Histograms of the Eddington-normalized beam power are
shown for each of the four samples studied here; the sample colors are 
as in Fig. 1. The mean value and standard deviations of the distributions are: 
$-2.74 \pm 0.63$ (576 LINERS from NB16); 
$-1.53 \pm 0.51$ (97 FRII sources from D16); 
$-2.50 \pm 1.11$ (80 compact radio sources from M03); and 
$-2.07 \pm 0.50$ (102 GBH from S15).}
		  \label{fig:newF2}
    \end{figure} 

\begin{figure}
    \centering
    \includegraphics[width=80mm]{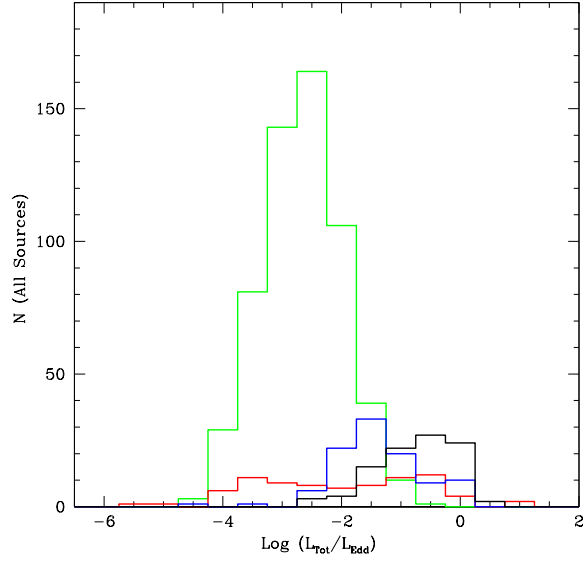}
\caption{Histograms of the Eddington-normalized total source power, 
where the total source power or luminosity is the sum of the bolometric 
luminosity and the beam power, are shown for each of the four samples 
studied here; the sample colors are as in Fig. 1. 
The mean value and standard deviations of the distributions are: 
$-2.66 \pm 0.67$ (576 LINERS from NB16); 
$-0.73 \pm 0.66$ (97 FRII sources from D16); 
$-2.01 \pm 1.40$ (80 compact radio sources from M03); and 
$-1.39 \pm 0.75$ (102 GBH from S15).}
		  \label{fig:newF3}
    \end{figure}

\begin{figure}
    \centering
    \includegraphics[width=80mm]{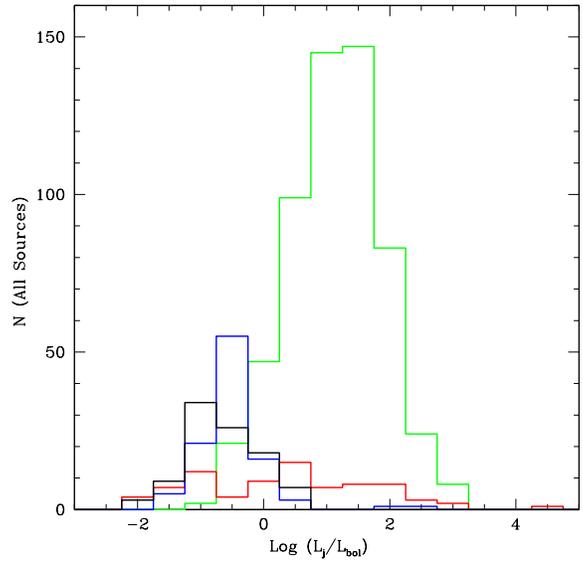}
\caption{Histograms of the radio of the beam power to the 
bolometric luminosity of the accretion disk, are shown for each of the 
four samples studied here; the sample colors are as in Fig. 1. 
The mean value and standard deviations of the distributions are: 
$1.14 \pm 0.73$ (576 LINERS from NB16); 
$-0.65 \pm 0.57$ (97 FRII sources from D16); 
$0.35 \pm 1.40$ (80 compact radio sources from M03); and 
$-0.50 \pm 0.56$ (102 GBH from S15).}
		  \label{fig:newF3}
    \end{figure}

\section{The Method}

The method introduced by D16 and expanded here is explained in section 3.1. 
A more intuitive approach to the derivation of the key equations is presented  
in section 3.2; the model assumptions are explained, and it is shown
that results obtained regarding the properties of the accretion disk
are independent of those obtained for black hole spins. 
The method, and the inputs to the method are all empirically based, 
and thus the results can be applied to study and constrain specific accretion 
disk and jet models, as explained in section 3.3. The uncertainties of 
the accretion disk properties and the black hole spin properties are also 
discussed in section 3.3.    

\subsection{The Method - Part 1}

All of the sources considered here 
fall on the FL of black hole activity, and the source properties  
indicate that the outflow is likely to be 
powered at least in part by the spin of the black hole, as discussed 
in section 2. Thus, the beam power
may be written as $L_j \propto \dot{m} ~M~f(j)$ $\propto B_p^2 ~M^2~f(j)$ 
(e.g. Blandford \& Znajek 1977; Blandford 1990; Meier 1999; 
Yuan \& Narayan 2014). Writing the bolometric luminosity of the disk as
$L_{bol} \propto \epsilon ~ \dot{m} ~M$,
and applying the empirical relationship given by eq. (1) leads to 
a derivation very similar to that presented in section 1.1, 
and it is easy to show that eq. (2) results. A second, more
intuitive derivation of eqs. (2) and (7) is discussed in Part 2 of this 
section.

As described in section 1.1, the maximum power of the accretion 
disk, $L_{bol}(max)$,  and outflow, $L_j (max)$, are parameterized
by the quantities $g_{bol}$ and $g_j$, respectively: 
$L_{bol}(max) = g_{bol} L_{Edd}$ and $L_j(max) = g_j L_{Edd}$.
The values of the normalization factors can be
estimated empirically (see D16 and D18), and can also be guided by 
theoretical considerations. 
The Eddington magnetic field strength, $B_{Edd}$, of a black hole system is 
$B^2_{Edd} \propto M^{-1}$ (e.g. Rees 1984; Blandford 1990; Dermer, 
Finke, \& Menon 2008), and, of course, $L_{Edd} \propto M$.  
(The Eddington magnetic field $B_{Edd}$ will have an
pressure similar to that of a radiation field with the Eddington luminosity.)
Thus, the beam power has the form 
\begin{equation}
\left({L_j \over L_{Edd}} \right) = g_j \left({B \over B_{Edd}} 
\right)^2 \left({f(j) \over f_{max}}\right),
\end{equation}
where $B$ is the total accretion disk magnetic field strength. 
If the energy density of the poloidal component of the magnetic field, $B_p$, 
is about $1/3$ of the total energy density of the field, 
then the maximum possible value of $g_j$ is 1/3. Many factors 
related to empirical and theoretical considerations can 
affect the value of $g_j$, as discussed in section 4.   

Combining eqs. (2) and (6), we obtain
\begin{equation}
\left({B \over B_{Edd}} \right)^2 = \left({L_{bol} \over g_{bol} L_{Edd}}\right)^{A}~.
\end{equation}  
Thus, for sources that fall on the FL, the slope of the FL, $A$, 
for that sample can be substituted into eq. (7) to obtain an empirical  
determination of the magnetic field strength $B$ in Eddington units. 
Note that a comparison of the equations described above indicates that 
$(B/B_{Edd})^2$ is identified with the dimensionless mass accretion 
rate, $\dot{m}$, $\dot{m} \propto (B/B_{Edd})^2$, with a constant of 
proportionality of order unity. The value of $A$ may be obtained by a direct fit to eq. (1),
as was done by D16 and D18, or by using the relationship 
between $A$ and best fit FP parameters included in 
eq. (3): $A= a/(a+b)$ (D18). 

In addition, the total magnetic field strength $B$ 
can then be obtained from the equation
\begin{equation}
\left({B \over 10^4 G} \right) = 
\left({B \over B_{Edd}} \right) \left({\kappa^2_B \over M_8}\right)^{1/2} 
= \left({L_{bol} \over g_{bol} L_{Edd}}\right)^{A/2} 
\left({\kappa^2_B \over M_8}\right)^{1/2} ~,
\end{equation}
where the Eddington magnetic field strength in units of $10^4$ G is 
$B_{Edd,4} \equiv \kappa_B M_8^{-1/2}$, where $\kappa_B \simeq 6$ (e.g. 
Rees 1984; Blandford 1990; Dermer, Finke, \& Menon 2008). 
 
\subsection{The Method - Part 2}

There is a second way to derive eqs. (2) and (7), the key equations
used to empirically quantify the properties of the accretion disk and 
black hole spin. This will illustrate 
the independence of the Eddington-normalized field strength $(B/B_{Edd})$
and normalized spin function $(f(j)/f_{max})$ obtained with the method
applied here. The key assumption is that the outflow is powered at
least in part by the spin of the black hole, as in the Blandford \& 
Znajek mechanism (1977) and the closely related 
Meier mechanism (1999), both of which have
the functional form given by eq. (6). 

Eq. (6) shows that the normalized beam power $[L_j/(g_j L_{Edd})]$ 
is separable into two independent parts: $(B/B_{Edd})^2$ and $(f(j)/f_{max})$. 
The first part, $(B/B_{Edd})^2$, depends  upon the properties of the 
accretion disk, since the magnetic field $B$ that threads the ergosphere 
of the black hole is anchored in the accretion disk. 
The second part, $(f(j)/f_{max})$, depends only 
upon the spin of the black hole, and is normalized by the maximum value
of the spin function. 

Thus, the assumptions are: that the
collimated outflow is powered at least in part by the spin of the black hole,
which leads to eq. (6); that the strength of the poloidal component 
of the magnetic field that is threading the black hole and is 
anchored in the accretion disk is independent
of the black hole spin; and that the spin of the black hole is 
independent of the conditions in the accretion disk. Thus, it is 
assumed that $(B/B_{Edd})$ and $(f(j)/f_{max})$ are independent. (Note
that the poloidal component of the magnetic field is some fraction 
of the full field strength $B$, as discussed in sections 3.1 and 4.)

Next, we note that four samples, including a sample of 
powerful extended radio sources and three samples of compact radio 
sources that fall on the FP, have an empirically determined relationship 
of the form $[L_j/L_{Edd}] \propto [L_{bol}/(g_{bol} L_{Edd})]^A$, which follows 
from eq. (1) with the disk normalization factor included.  
This empirical relationship only involves the properties of the accretion disk, 
indicated by the right hand side of the equation, and the properties of the 
outflow, indicated by the left hand side of the equation; the spin is
not included in this empirically determined relationship.  
The term $[L_{bol}/(g_{bol} L_{Edd})]^A$ is identified 
with the term $(B/B_{Edd})^2$ on the right hand side of eq. (6); that is, 
$(B/B_{Edd})^2 = [L_{bol}/(g_{bol} L_{Edd})]^A$, which is eq. (7). This is
valid because it is assumed that the spin of the black hole is independent of
the accretion disk properties. Substituting eq. (7) into eq. (6) leads to 
eq. (2). 

This alternative derivation makes it clear that the 
field strength obtained with eqs. (7) or (8) is independent of the spin 
obtained with eq. (2) given the assumption that the outflow is described by
eq. (6), and thus is separable into two independent functions, 
$(B/B_{Edd})^2$ and $(f(j)/f_{max})$. When these functions are obtained
empirically, it is easy to check and confirm that there is no covariance 
between them, which indeed turns out to be the case. 

\subsection{The Method - Part 3}

 The ``outflow method'' of estimating black hole spins and 
accretion disk magnetic field strengths proposed by D16 and extended  
here can only be applied to certain categories of sources, 
as described in section 1.1. It is an empirically based method,
and thus is independent of a specific accretion 
disk emission model, and is independent of a specific jet model relating beam
power to compact radio luminosity. The key assumptions are described in
section 3.2. 

The sources selected for study follow the FL of black hole activity, given 
by eq. (1) (see D18 for other forms of this equation). 
This includes sources that were placed directly on the FL,
such as the FRII sources, and FP sources, as described in section 1.1.   
This means that the sources are empirically well described by only 
a few parameters. 
As it turns out, the samples studied are described by similar 
parameter values over a very large range of parameter space 
(described here by the parameter $A$), and thus are likely to be
controlled by similar physical processes, as has been recognized for 
FP sources for quite some time (e.g. Corbel et al. 2003; Gallo et al. 2003;
M03; Falcke et al. 2004; Kording et al. 2006; Gultekin et al. 2009;
Bonchi et al. 2013; van Velzen \& Falcke 2013; S15; NB16). 

Thus, these sources are likely
to have a special set of characteristics. For example, it is likely 
that the accretion disk magnetic field strength $B$ is either in equipartition
with the gas and radiation pressure $P$ of the accretion disk or is 
some constant fraction of this pressure, so $P \propto B^2$ 
(e.g. Balbus and Hawley 1991, 1998; Hawley \& Balbus 1991), that
the magnetic field is threading the ergosphere of the black hole 
(a requirement for the spin powered outflow model), 
and that the black holes have substantial spin.
Substantial spin is required because the outflow beam power is 
proportional to the square of the black hole spin, so black holes 
with low spin will produce weak outflows via this process, as 
discussed in section 6.   

This means that the empirical results, once obtained, can be applied to study 
and constrain detailed outflow and accretion disk models since these 
models are not an input to any of the results presented here. That is,
the gas and radiation pressure of the disk can be estimated from 
the quantity $B^2$ obtained using eq. (8), and the relationship between
the bolometric disk luminosity, the disk pressure $P$, and the black hole
mass can then be studied using eq. (7) or (8). For example, eq. (7)
indicates that $(B/B_{Edd})^2 \simeq \dot{m} \simeq
(L_{bol}/L_{Edd})^{1/2}$ for $g_{bol} = 1$ and $A \simeq 0.5$ (see
Tables 1 and 2 from D18), which is consistent with expectations of 
radiatively inefficient accretion disk models (see, for example,
eq. (4) of Ho 2009), though these models may
have to be extended or modified to allow values as high as $L_{bol}
\sim L_{Edd}$ (e.g. Yuan \& Narayan 2014). 
Similarly, the beam power for the FP sources 
is obtained by D18, as described in section 1.1,  
without specifying a detailed physical model for the jet or for 
the relationship between the beam power and the 
radio luminosity of the compact radio source. 
The beam power is empirically determined using eq. (4). Thus, detailed 
physical jet models that reproduce the observed radio emission given the 
beam power could be explored, such as the models discussed by  
Heinz \& Sunyaev (2003), M03, Heinz (2004), Yuan \& Cui (2005), Yuan et al. (2005), and 
Li et al. (2008), for example.  

The uncertainties of the input parameters $L_{bol}$, $L_j$, and $L_{Edd}$ 
will impact the uncertainties of $(B/B_{Edd})$, $(B/10^4 G)$, and 
$\sqrt{(f(j)/f_{max})}$. To compute these uncertainties, we adopt 
a value of $A \simeq 0.45$ for all sources and use 
the shorthand $B_4 \equiv B/(10^4 G)$ and $F \equiv \sqrt{(f(j)/f_{max})}$. 
Following the discussion provided at the end of section 2, and
considering eq. (7), it is easy to 
show that $\delta \rm{Log}(B/B_{Edd}) = (A/2) ~[(\delta \rm{Log}(L_{bol}))^2 
+ (\delta \rm{Log}(L_{Edd}))^2]^{1/2}$, so for FRII AGN with 
$\delta \rm{Log}(L_{Edd}) \simeq 0.3$ and $0.4$, we obtain 
$\delta \rm{Log}(B/B_{Edd}) \simeq 0.11$ and $0.12$, respectively. 
It is assumed that $g_{bol} = 1$ for all sources, that is, the maximum possible 
luminosity of the disk is equal to the Eddington luminosity, which
is consistent with the samples studied here, as discussed by D16 and
D18. Adopting values of $\delta \rm{Log}(L_{Edd}) \simeq 0.3$ to 0.4 
for the LINERS, we obtain $\delta \rm{Log}(B/B_{Edd}) \simeq 0.09$ and 0.11,
respectively. For FP sources with $\delta \rm{Log}(L_{Edd}) \simeq 0.5$, 
such as the M03 sources, $\delta \rm{Log}(B/B_{Edd}) \simeq 0.13$. For the GBH, 
$\delta \rm{Log}(B/B_{Edd}) \simeq 0.06$. 
For the magnetic field strength in physical units, eq. (8) indicates
$\delta \rm{Log}(B_4) = [(A/2)^2 ~(\delta \rm{Log}(L_{bol}))^2 
+ [(A+1)/2]^2 ~(\delta \rm{Log}(L_{Edd}))^2]^{1/2}$, so $\delta \rm{Log}(B_4) \simeq 
0.23$, $0.22$, and $0.37$ for the FRII AGN, LINERS, and M03 sources
respectively, adopting $\delta \rm{Log}(L_{Edd}) \simeq 0.3$ for the 
LINERS and FRII AGN and $\delta \rm{Log}(L_{Edd}) \simeq 0.5$ for 
the M03 sources.  For a value of $\delta \rm{Log}(L_{Edd}) \simeq 0.4$
for the FRII AGN and LINERS, these values become 0.30 for both FRII AGN and
LINERS. For the GBH, $\delta \rm{Log}(B_4) \simeq 0.06$. 
Eq. (2) indicates that  
$\delta \rm{Log}(F) = 0.5[(\delta \rm{Log}(L_j))^2 
+ (\delta \rm{Log}(g_j))^2 + (A-1)^2(\delta \rm{Log}(L_{Edd}))^2+
A^2(\delta \rm{Log}(L_{bol}))^2]^{1/2}$. Leaving aside $(\delta g_j/g_j)$ for 
the present time, for the FRII AGN, $\delta \rm{Log}(F) \simeq 0.14$ 
and 0.15 for $\delta \rm{Log}(L_{Edd}) \simeq 0.3$ and 0.4, respectively. 
For the LINERS, $\delta \rm{Log}(F) \simeq 0.13$ 
and 0.15 for $\delta \rm{Log}(L_{Edd}) \simeq 0.3$ and 0.4, respectively. 
For the M03 sources, $\delta \rm{Log}(F) \simeq 0.19$. 
For the GBH, $\delta \rm{Log}(F) \simeq 0.07$. 
These estimated uncertainties per source are included in parenthesis in columns
(4), (7), and (8) in the top part of Table 1. 

\begin{table*}
\begin{minipage}{165mm}
%\scriptsize
\caption{Mean Value and Standard Deviation of Histograms and Values for Select Individual Sources.
{%
\footnote{Obtained for $g_{bol} = 1$ and $g_j = 0.1$ for all sources. 
The estimated uncertainty per source is included in brackets following the
standard deviation in columns (4), (7), and (8) 
for the samples listed in the top part of the table, as discussed in section 3.3.
The bottom part of the table includes entries for three individual GBH, which have 
multiple observations per source, one additional GBH, and several 
individual AGN.}}}   % title of Table
\label{tab:comp}        % is used to refer this table in the text
%\centering                          % used for centering table
\begin{tabular}{llllllll}   % centered columns (4 columns)
\hline\hline                    % inserts double horizontal lines
(1)&(2)&(3)&(4)&(5)&(6)&(7)&(8)\\
Sample&Type&N
{%
\footnote{N is the number of sources for the AGN and the number of 
measurements for the GBH.}}
&${\rm{Log}}\sqrt{(f(j)/f_{max})}$
&$j$&$j_{pub}$ 
(ref){%
\footnote{Published spin values; the citations are: (1) Miller et al. (2009); 
(2) Garcia et al. 2015; (3) Vasudevan et al. (2016); (4) Patrick et al. (2012); 
and (5) Walton et al. (2013).}}
&$\rm{Log}(B/B_{Edd})$&$\rm{Log}(B/10^4 G)$\\
\hline	
NB16(1){%
\footnote{Parameters are from the first line of Table 3 of NB16.}}
&AGN&576&$ -0.04 \pm 0.24 (0.14)$&$0.93 \pm 0.10$&&
$	-0.83	\pm	0.21 (0.10)	$&$	-0.09	\pm	0.39 (0.22)$\\
D16&AGN&97	&$-0.07 \pm 0.19 (0.15)$&$0.93 \pm 0.11$  &&$	-0.19	\pm	0.17 (0.12)	$&$	0.03	\pm	0.23 (0.23)	$\\
M03&AGN&80	&$-0.17 \pm 0.36 (0.19)$&$0.81 \pm 0.20$&&$	-0.58	\pm	0.42 (0.13)	$&$	0.27	\pm	0.66 (0.37)	$\\
S15&GBH&102	&$-0.17 \pm 0.10 (0.07)$  &$0.92 \pm 0.09$&&$	-0.37	\pm	0.23 (0.06)	$&$	4.00	\pm	0.24 (0.06)	$\\
\hline
\hline
GX	339-4& GBH&76&$-0.17 \pm 0.06$&$0.92 \pm 0.06$&$0.94 \pm 0.02$ (1)&
$-0.32\pm0.18	$&$4.07	\pm	0.18	$\\
GX	339-4& &&&&$0.95_{-0.05}^{+0.03}$ (2)&&\\
V404 Cyg& GBH&20	&$-0.10 \pm 0.06$&$0.97 \pm 0.02$&&$	-0.44	\pm	0.24	$&$	3.84	\pm	0.24	$\\
XTE J1118
{%
\footnote{The full source name is XTE J1118+480. }}
& GBH&5&$-0.43 \pm 0.01$&$0.66 \pm 0.02$&&
$	-0.54	\pm	0.01	$&$	3.80	\pm	0.01	$\\
AO6200~{%
\footnote{The uncertainties 
for this source and the other individual sources listed here are
estimated as described in section 3.3. }}
& GBH&1	&$-0.08 \pm 0.07$
&$0.98 \pm 0.07$&&$-1.62 \pm 0.06$&$2.76 \pm 0.06$\\
Sgr	$A^*~$
& AGN&1&$-0.17 \pm 0.19$&$0.93 \pm 0.15$&&$-2.09 \pm 0.13$&$-0.61 \pm 0.37$\\
M87
{%
\footnote{Also referred to as NGC 4486.}}
& AGN&1&$0.13 \pm 0.19$&$1.00 \pm 0.15$&&$-1.19 \pm 0.13$&$-1.16 \pm 0.37$\\
Ark 564&  AGN& 1&$0.06 \pm 0.19$&$1.00 \pm 0.15$ &$0.96_{-0.11}^{+0.01}$ (3)&$	0.17 \pm 0.13$ &$1.93 \pm 0.37$\\
Mrk 335&  AGN&1&$-0.29 \pm 0.19$&$ 0.81\pm 0.15$ &$> 0.91$ (3)  &$-0.15 \pm 0.13$&$1.06\pm 0.37$\\
Mrk 335&&&&&$0.70_{-0.01}^{+0.12}$ (4)&\\
Mrk 335&&&&&$0.83_{-0.13}^{+0.10}$ (5)&\\
NGC 1365&  AGN&1& $0.53 \pm 0.19$&$	1.00\pm 0.15$&$> 0.84$ (3)&$-0.64 \pm 0.13$&$0.70\pm 0.37$\\
NGC 4051&  AGN& 1&$-0.02 \pm 0.19$&$1.00 \pm 0.15$&$ > 0.99$ (3)  &$ -0.34 \pm 0.13$&$	1.30\pm 0.37$\\
NGC 4151&  AGN&1&$-0.27 \pm 0.19$& $0.84 \pm 0.15$&$>0.9$ (3) &$-0.35 \pm 0.13$&$ 	0.60\pm 0.37$\\
3C 120&  AGN&1&$0.59 \pm 0.19$ &$1.00\pm 0.15$ &$>0.95$ (3) &$-0.13 \pm 0.13$& $0.78\pm 0.37$\\
\hline 
\end{tabular}
\end{minipage}
\end{table*}

\begin{figure}
    \centering
    \includegraphics[width=80mm]{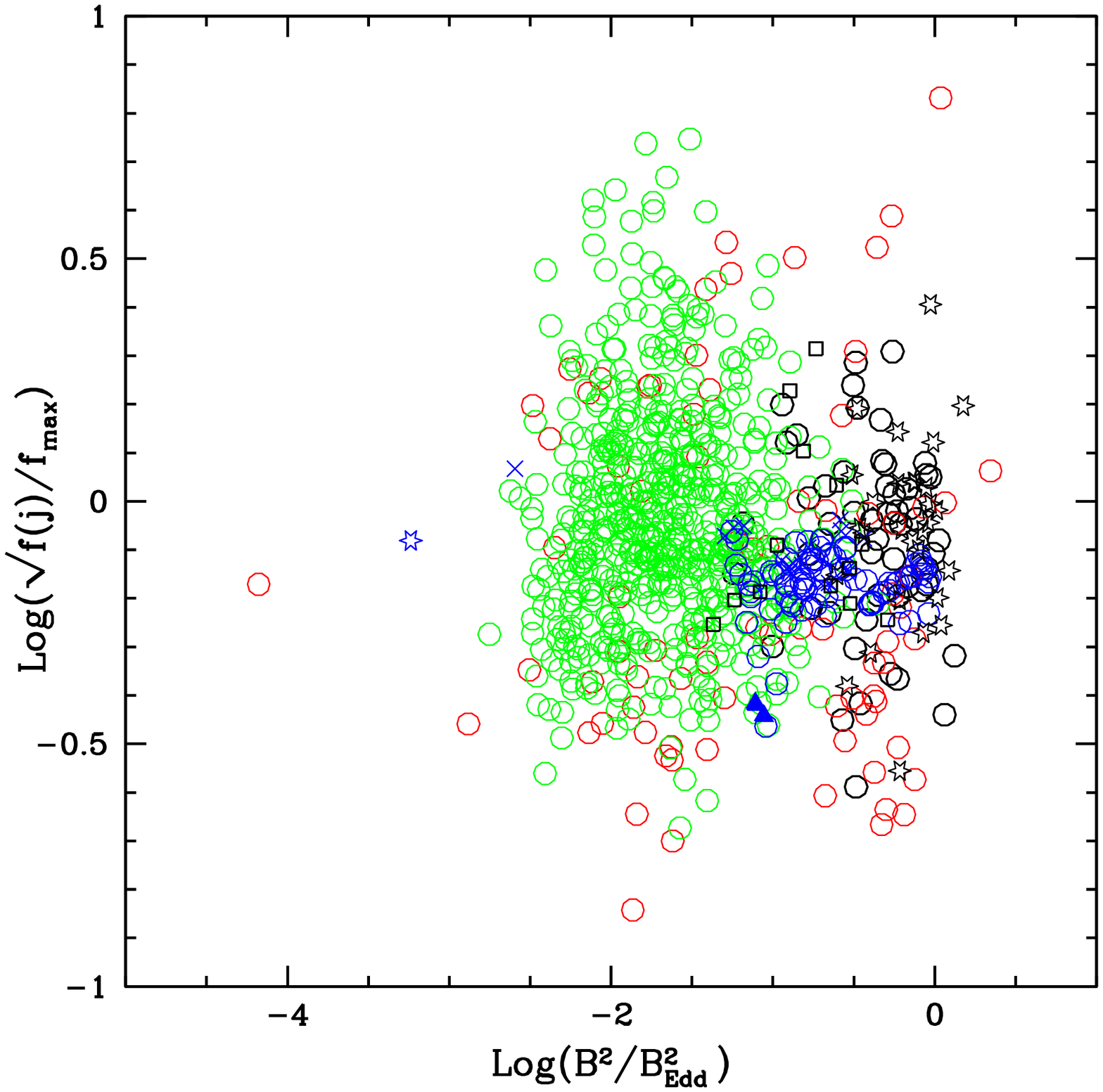}
\caption{The square root of the black hole spin function is shown versus the 
square of the accretion disk magnetic field strength in Eddington units.  
Here and throughout the paper: the 576 LINERS from NB16
are shown as green circles; the 97 FRII sources from D16 
are shown in black (circles represent 55 
high excitation radio galaxies, stars represent 29 radio loud quasars,
and squares represent 13 low 
excitation radio galaxies); the 80 compact radio sources
from M03 are shown as red circles; and the 102 GBH from 
S15 are shown in blue (open circles represent GX 339-4,
crosses represent V404 Cyg, solid triangles represent XTE 
J1118+480, and an open star represents AO 6200). 
Values of $g_{bol} = 1$ and $g_j = 0.1$ have been adopted for all sources.
The source with a value of $\rm{Log}(B^2/B_{Edd}^2) \simeq -4$ is Sgr $A^*$. 
Fits show that there is no covariance between the quantities 
$\rm{Log}(\sqrt{f(j)/f_{max}})$ and $\rm{Log}(B/B_{Edd})^2$. }
		  \label{fig:F1}
    \end{figure} 

\section{Results}

The spin function, $f(j)/f_{max}$, is obtained using eq. (2), the 
accretion disk magnetic field strength in Eddington units, 
$(B/B_{Edd})$, is obtained using eq. (7), and the magnetic field strength
is obtained in physical units using eq. (8) for each 
source. Once obtained, it turns out that there is no covariance between 
the spin function and either of the field strengths;  
they are independent, as expected (see section 3.2). Results for each sample
and some specific sources are summarized in Table 1. 

Source parameters are obtained as described in sections 1, 2, and 3. 
These parameters including name, type, redshift or distance, Eddington
luminosity, beam power, bolometric luminosity, square root
of the normalized spin function, black hole spin, Eddington normalized
magnetic field strength, and magnetic field strength in units of $10^4$ G are
listed in Tables 2 - 5 for each of the samples studied; all luminosities 
are in erg/s. Results are also displayed in Figs. 8 - 14, and 
the mean values and dispersions of the histograms are summarized
in Table 1. This table also includes information on each of the four
GBH systems, Sgr A*, M87, and 6 AGN systems that have spin values 
obtained with the X-ray reflection method; systems
described by Reynolds (2019) as having ``robust'' spin values 
that overlap with the sources studied here are included in Table 1.   
Three W galaxies from Grimes et al. (2004) are also included at the end 
of Table 4, and a second observation of AO6200 is discussed in section 
5. 
 
The values of $g_{bol}$ and $g_j$ can be inferred empirically.
The empirical results presented by D16 and D18 suggest that   
$g_{bol} \simeq 1$ and $g_j \simeq 0.1$; these values are adopted here. 
It is easy to scale the results for other values of these parameters. 
As discussed in section 3.1, theoretical considerations suggest a maximum 
value of $g_j$ of about 1/3, and other factors 
can also affect the value of $g_j$. Empirically, there could be a 
normalization offset of the value of $L_j$ from the true value, 
and this would be
absorbed into the value of $g_j$ when this is determined by 
the maximum measured value of $L_j/L_{Edd}$. In addition, there may be variations
in the ratio of $(B_p/B)^2$, leading to variations in the parameter $g_j$.
Interestingly, the values of $g_j$ indicated by the Meier (1999) model
are $g_j \simeq 0.3$ for $(B_p/B)^2 \simeq 1$, and $g_j \simeq 0.1$
for $(B_p/B)^2 \simeq 1/3$. This can be seen by considering eq. (1) of
Daly (2011), noting that the magnetic field in that equation is the 
poloidal component only, replacing $j$ with $\sqrt{f(j)/f_{max}}$, 
and writing the equation in the form of eq. (6). 
Thus, variations in the fraction of the total
magnetic field that is in the poloidal component cause variations in 
$g_j$, which would affect determinations of $\sqrt{f(j)/f_{max}}$ 
and hence $j$. As $g_j$ increases from the value of 0.1 adopted here,
the value of $\sqrt{f(j)/f_{max}}$ and hence $j$ will decrease from the values
listed here. For $1/3 \leq (B_p/B)^2 \leq 1$, this limits the range of
values of $g_j$ to (0.1 - 0.3) in the context of the Meier (1999) model. 

The spin function $f(j)/f_{max}$ is obtained using eq. (2) and the 
Eddington normalized magnetic field strength $(B/B_{Edd})$ 
is obtained using eq. (7) applying the 
value of $A$ listed in column 3 of Table 2 from D18 for each of the 
four samples considered here. 
Values of the spin function and Eddington normalized magnetic field
are illustrated in Fig. 8. The cutoff of the sources at 
$(B/B_{Edd})^2 \simeq 1$ reflects the cutoff of sources at $L_{bol} \simeq L_{Edd}$,
as discussed by D16 and D18. Statistical fluctuations and 
measurement uncertainties allow sources to have values of
$f(j)/f_{max} > 1$. 

\begin{figure}
    \centering
    \includegraphics[width=80mm]{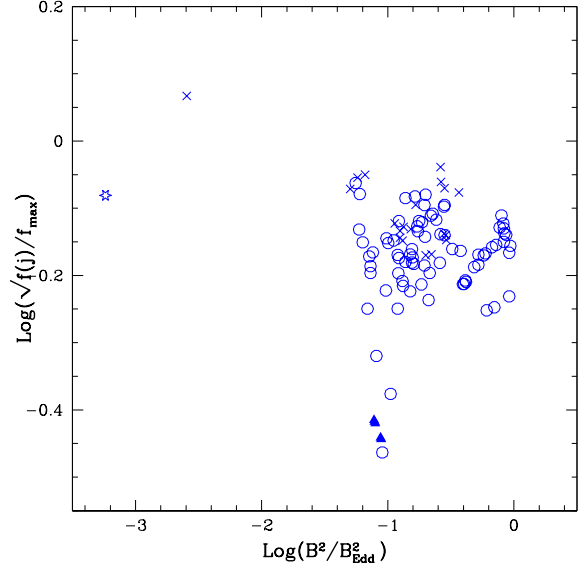}
\caption{As in Fig. 8 with only results for GBH shown. There are 76 
measurements for the source GX 339-4 shown as open circles; 20 
measurements for the source V404 Cyg shown as crosses; 
5 measurements for the source XTE J1118+480 shown as solid triangles; and
1 measurement for the source AO 6200 shown as an open star (a second
observation of AO 6200 is discussed in section 5). }
		  \label{fig:2}
    \end{figure} 

The GBH systems have multiple simultaneous radio and X-ray measurements; the 
values used here are from S15, who obtained them from Corbel et al. (2013)
for GX 339-4; Corbel, K{\"o}rding, \& Kaaret (2008) for  V404 Cyg;
Wagner et al. (2001) for XTE J1118+480; and Gallo et al. (2006) for 
AO6200. Each set of measurements yields a 
value of the spin function and magnetic field strength, as shown in Figs.
8, 9, and 12, and summarized in Table 1. 
As can be seen in Fig. 9, the measurements for each source tend to 
cluster around a specific value of the spin function, and, in general, 
the spin functions tend to have a smaller dispersion than the magnetic 
field strengths; these differences
are evident in the mean values and dispersions listed in Table 1.  
It is interesting that AO6200 has 
a much lower magnetic field strength but a similar spin value compared
with the other GBH. (Another observation of this source from 1975-76 
is discussed in section 5, and indicates a similar spin value but a
significantly higher field strength.)  
Having multiple observations per source 
allows a comparison of spin and magnetic field strength values  
for a particular GBH indicated by measurements obtained at different times. 
The spin indicator $\sqrt{(f(j)/f_{max})}$ of GX 339-4 and V404 Cyg  
have a very small dispersion, while the dispersions of the magnetic field
strengths are larger; recall that $\sqrt{(f(j)/f_{max})} \propto j$ to first
order in $j$, so this is the quantity studied here as described in section 1.1.
As will be discussed in section 5, this suggests that the spin
of the GBH remains constant (as expected), but the accretion disk magnetic field
strength is time variable, and varies as the accretion rate changes. 
This impacts the beam power of the outflow, since this is regulated in part
by the magnetic field strength of the accretion disk. 

\begin{figure}
    \centering
    \includegraphics[width=80mm]{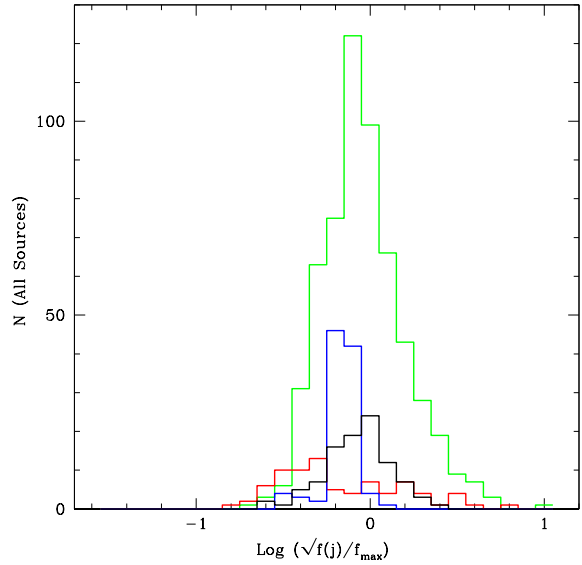}	 
\caption{Histograms of the square root of the spin function shown in Fig. 10 
for each of the four samples considered. As in all of the figures, the 
the 576 LINERS from NB16
are shown in green; the 97 FRII sources from D16 
are shown in black; the 80 compact radio sources
from M03 are shown in red; and the 102 GBH from 
S15 are shown in blue. The results are summarized in Table 1.}
		  \label{fig:F3}
    \end{figure} 

The histograms of the black hole spin function are shown for each 
sample separately in Fig. 10, and the mean values and dispersions of
the histograms are listed in Table 1. All of the source types, 
including LINERS (NB16), powerful extended (FRII) radio sources (D16), 
compact radio sources that are AGN (M03), and GBH (S15) have similar 
spin function mean values. This indicates that source type is not related to 
black hole spin for the sources with collimated outflows studied here. 
And, this indicates that black hole spin does not determine
AGN type for the types of sources considered here. 

This is consistent with the result obtained by D16 for FRII sources. 
D16 found that   
subtypes of FRII sources, including high excitation radio galaxies (HEG), 
low excitation radio galaxies (LEG), and radio loud quasars (RLQ), had similar 
spin functions and spin function distributions. The results obtained here
indicate that the mean value and dispersion of all quantities for 
each of the subsamples,  including 55 HEG, 
29 RLQ, and 13 LEG, are very similar to those of
the full sample of 97 sources with the exception of $\rm{Log}(B/B_{eq})$, 
which has values of $-0.20 \pm 0.16$ for 55 HEG, $-0.09 \pm 0.10$ for 
29 RLQ, and $-0.39 \pm 0.16$ for 13 LEG.  That is, different
FRII types are not distinguishable by mean value or dispersion 
of spin function, spin value, or magnetic field strength in physical
units, indicating that black hole spin does not determine 
AGN type for these three types of classical double (FRII) sources. 

\begin{figure}
    \centering
    \includegraphics[width=80mm]{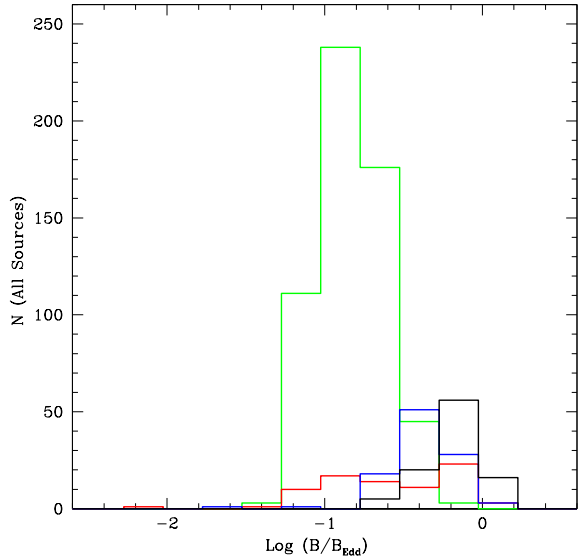}	 
\caption{Histograms of the Eddington normalized accretion disk 
magnetic field strengths shown in Fig. 8 for each of the four samples 
considered. The colors are as in Fig. 1, and the results are summarized 
in Table 1. }
		  \label{fig:F4}
    \end{figure} 

Histograms of the accretion disk magnetic field strength in units of the
Eddington field strength are shown Fig. 11, and are obtained 
as described above. The mean values and dispersions of the field strength 
are listed in Table 1 and mirror those of the bolometric disk luminosity 
in Eddington units (see eq. 7). There are a broad range of values of 
the Eddington normalized disk magnetic field strength, as expected
since this tracks the Eddington normalized bolometric luminosity of the disk.
Insofar as the Eddington normalized disk luminosity varies with AGN
type (e.g. Ho 2009), so does the Eddington normalized magnetic field
strength. As noted in section 3.1, $\dot{m} \simeq (B/B_{Edd})^2$, and 
Fig. 8 indicates that the dimensionless mass accretion rate spans
about 2.5 orders of magnitude, ranging from about $3 \times 10^{-3}$
to 1, with sharp cut-offs at both the low and high end. The cutoff at
low values of $(L_{bol}/L_{Edd})$ is likely due to a survey flux
limit (e.g. NB16 for the LINERS), while that at the high end occurs at 
$L_{bol} \simeq L_{Edd}$.  

The black hole spin may be obtained from the spin function once
a particular outflow model is adopted, and are shown in Fig. 12 for
the values of $g_j$ and $g_{bol}$ adopted as described above. 
As discussed in section 1.1, numerical simulations suggest that  
the conversion from the spin function to 
spin can change depending upon the details of the model for the black hole 
system (e.g. Tchekhovskoy et al. 2010; Yuan \& Narayan 2014), 
so this may be a source of uncertainty. 
Here, the standard conversion from the spin function to black hole spin,  
$\sqrt{f(j)/f_{max}} = j (1+\sqrt{1-j^2})^{-1}$, is adopted. For 
${f(j)/f_{max}} \leq 1$, this implies 
\begin{equation}
j = {{2 \sqrt{f(j)/f_{max}}}  \over {f(j)/f_{max} + 1}}.
\end{equation}

\begin{figure}
    \centering
    \includegraphics[width=80mm]{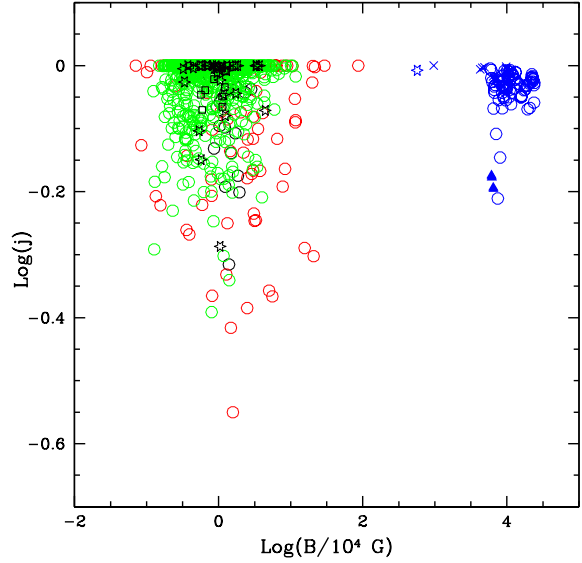}	 
\caption{Black hole spin is shown versus the accretion disk
magnetic field strength in units of $10^4$ G. The 
black hole spin is obtained from the spin function 
as described in the text. When the spin function 
$f(j)/f_{max} \geq 1$, the spin $j$ is set equal to one. 
The symbols are as in Fig. 8.}
		  \label{fig:F5}
    \end{figure} 

\begin{figure}
    \centering
    \includegraphics[width=80mm]{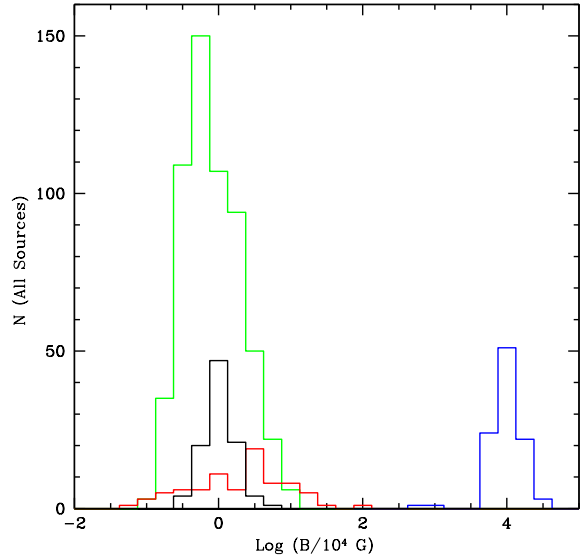}	 
\caption{Histograms of the accretion disk magnetic field
strengths shown in Fig. 12 for each of the four samples considered. 
The colors are as in Fig. 1, and the results are summarized in Table 1. }
		  \label{fig:F6}
    \end{figure} 

The black hole spin is obtained using eq. (9), and 
values of ${f(j)/f_{max}} > 1$ are set equal to one.
This causes asymmetries in the distribution of spin values when
many sources have values of ${f(j)/f_{max}} > 1$. 
Values of $j$ thus obtained are shown in Fig. 12, 
histograms of spin values are shown in Fig. 14; and 
values are listed in Tables 1-5. 

The accretion disk magnetic field strength in physical units is
obtained using eq. (8), and is illustrated in Fig. 12. It is easy 
to see that the accretion disk magnetic field strengths for AGN
are quite similar for all AGN types, and have a broad range
of values with a peak at about $10^4$ G. 
The accretion disk magnetic field strengths for the GBH measurements 
are all quite similar have values of about $10^8$ G, 
with the exception of the one measurement of AO6200 and one 
one measurement of V404 Cyg. 
Histograms of the accretion disk magnetic field strength in physical 
units for each sample are
shown in Fig. 13, and the values are summarized in Table 1. 

Histograms of the black hole spin are shown in Fig. 14 and summarized
in Table 1. As noted earlier, spin values of sources with spin functions 
greater than one indicate a spin value of one, as is evident in Fig. 14, 
and some of the spin distributions are clearly asymmetric. 

\section{Discussion}

Three key results are presented in section 4 related to black hole spin.
(1) Black hole spin functions and spin values 
obtained for all source types studied are relatively high. 
For the NB16, D16, and S15 samples, the mean spin values are about 0.9,
and for the M03 sample the mean spin value is about 0.8 (see Table 1). 
As noted earlier, the spin distributions become 
skewed when  $\sqrt{f(j)/f_{max}} > 1$,
since in this case the spin is set equal to one, so the mean value of the 
spin (listed in Table 1) is smaller than that indicated by the mean 
value of the spin function. (2) All of the sources types 
studied, including LINERS (NB16), FRII sources (D16), GBH (S15), and 
a compilation of compact radio sources (M03), have similar values of the spin
function $\sqrt{(f(j)/f_{max})}$ and black hole spin $j$. This suggests 
that AGN type is not determined by black hole spin 
for the types of sources studied here. This is consistent with the finding
of D16, that subtypes of FRII sources including
HEG, RLQ, and LEG have similar spin functions and spin values 
and thus their AGN type is not determined by black hole spin.  (3) For 
the GBH with numerous measurements per source, 
the dispersions of the spin indicators, including the spin 
function and spin value, are smaller than those of the magnetic field strengths.
The implications of these results will be discussed below. 

\begin{figure}
    \centering
    \includegraphics[width=80mm]{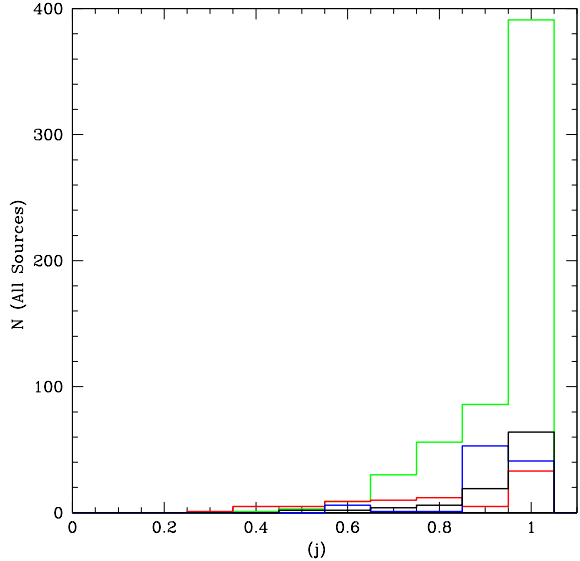}	 
\caption{Histograms of spin values for each of the four samples 
shown in Fig. 12. The colors are as in Fig 1. 
The results are summarized in Table 1. }
		  \label{fig:F7}
    \end{figure} 

Three key results are presented in section 4 related to accretion disk magnetic
field strengths. (1) There is a range of mean values of the Eddington
normalized magnetic field strengths, and these are related to source type.
This is not surprising since for AGN some source types are related to 
$(L_{bol}/L_{Edd})$ (e.g. Ho 2009). (2) The distributions of 
accretion disk magnetic field strengths  
in physical units are broad and similar for the AGN samples studied, 
and peak at about $10^4$ G. Thus, the accretion disk magnetic field strength
in physical units is not related to AGN type. The 
field strength distributions for the GBH are similar for most of 
the measurements
and peak at about $10^8$ G, with the exception of the one 
measurement of AO6200 and one measurement of V404 Cyg. 
(3) The dispersion of the magnetic field 
strengths, both in Eddington units and in physical units,   
is larger than that of the spin indicators for the GBH sample 
and for the two individual GBH with numerous observations per source.   

Point (3) from each paragraph above can be understood in the 
following way. Over the course of the numerous simultaneous radio
and X-ray observations of the GBH,  
the spin of the black hole remains constant, but the 
accretion disk magnetic field strength varies with 
the accretion rate. This leads to the field strengths having a
larger dispersion than the spin indicators. 
The variation of the accretion disk magnetic field strength 
impacts the beam power of the outflow, since the beam power  
is regulated in part by the magnetic field strength of the disk,
but it does not impact the black hole spin. The fact that there
are numerous simultaneous radio and X-ray observations for the 
GBH allow the changes in the accretion rate and resulting changes
in the beam power to be studied. 

Black hole spins obtained here can be compared with those 
obtained independently. GX 339-4 has published spin values 
of $0.94 \pm 0.02$ (Miller et al. 2009) and $0.95_{-0.05}^{+0.03}$ 
(Garcia et al. 2015), and the value obtained here is 
$0.92 \pm 0.06$ (see Table 1). Thus, the value obtained here is in
good agreement with that indicated by a well established and 
independent method. 
For AGN there is good agreement between the spin values obtained
here and those obtained with the X-ray reflection method 
(see Table 1). The spin values obtained with the X-ray reflection 
method are from Patrick et al. (2012), Walton et al. (2013), 
and the compilation of Vasudevan et al. (2016). Only AGN
with X-ray reflection spin measurements described by Reynolds (2019)
as being ``robust'' were compared with spin determination obtained here. 
The uncertainties of the spin values for individual AGN are estimated 
as described in section 3.3. 

The value of the spin for AO6200 obtained here is relatively high. 
Gou et al. (2010) used a single high-quality X-ray spectrum from 1975 and 
obtained a low value for the spin of this source by fitting the
spectrum with a detailed model of the accretion disk. The X-ray 
data suggest a value of $L_{bol}/L_{Edd} \simeq 0.11$ (Guo et al. 2010). 
Kuulkers et al. (1999) report radio observations obtained at about the 
same time as
the X-ray outburst. We consider the 1.4 GHz data point with a flux
density of about 0.3 Jy obtained by Owen et al. (1976) and listed
in Table 1 of Kuulkers et al. (1999) because
it is comparable in frequency to the FP constructed by NB16, to which
all of the compact radio sources studied here are scaled. Converting
this flux density to a flux by multiplying by 1.4 GHz, and using
the distance to the source provided by Gou et al. (2010), a value
of $\rm{Log}(L_j) \simeq 37.23$ is obtained from the radio luminosity
of $\rm{Log}(L_R) \simeq 29.74$ using the values of C and D from 
Table 1 of D18 following the procedure used for all of the GBH discussed
here and in D18. A value of 
$\rm{Log}(B/B_{Edd}) \simeq -0.23$ is indicated by $L_{bol}/L_{Edd} \simeq 0.11$,
obtained with eq. (7) for $A \simeq 0.47$ applicable for GBH (D18). 
Combining the value of $\rm{Log}(L_j) \simeq 37.23$ with the 
the value of $L_{Edd}$ listed in Table 2 to obtain $L_j/L_{Edd}$, 
and assuming the standard values of  
$g_{bol} = 1$ and $g_j = 0.1$, a value of 
$\rm{Log}\sqrt{(f(j)/f_{max})} \simeq -0.11$ is obtained with eq. (2), 
and a spin of about 0.97 is obtained with eq. (9). 
Interestingly, this is similar to the spin value of $0.98$ obtained 
here (see Table 1).
Note, that the field strength listed in Table 2 is significantly different,
and is $\rm{Log}(B/B_{Edd}) \simeq -1.62$. Thus, even though this field strength
is significantly lower than that indicated during the outburst, the 
combination of contemporaneously obtained radio and X-ray data
yield a value of the spin that is consistent. 
Thus, the empirical results indicate that the magnetic field strength of the 
accretion disk is variable, but the spin of the black hole
remains roughly constant, as expected in the context of the outflow-based 
method applied here (e.g. eq. 6). 

Accretion disk magnetic field strengths have been estimated for GBH
in the context of an outflow model for these system by Piotrovich et al. 
(2015a), who report values of $\sim 10^8$ G for these systems. Accretion
disk magnetic field strengths have been estimated for AGN in the 
context of a detail accretion disk model, and values of $\sim 10^4$ G
are reported (e.g. Mikhailov, Gnedin, \& Belonovsky 2015; Piotrovich et al. 
2015b). Thus, the accretion disk magnetic field strengths obtained here 
for GBH and AGN are similar to values indicated by other methods.

Dispersions of the distributions of empirically determined
quantities for each of the samples listed in Table 1 can be compared
with the uncertainty per source estimated based upon the uncertainties
of the input parameters, as discussed in sections 2 and 3.3. 
The dispersions of the full population are significantly larger than 
the measurement uncertainty per source divided by the square root of
the number of sources per sample. 
Thus, the results are consistent with the sources having a broad 
distribution of each of the quantities studied here.

\section{Summary}

The ``outflow method'' of determining black hole spin functions and 
spins proposed and applied to a sample of FRII sources by 
D16 is applied to the samples studied by D18. This requires a value of 
the outflow beam power $L_j$ for each source. Beam powers are obtained for FP 
sources using the method proposed and applied by D18 and explained in section 1.1, 
and values for the FRII sources are obtained using the strong shock method.  
Values of $L_{bol}$ and $L_{Edd}$ are obtained using standard
methods. Eq. (2) is used to solve for the spin function using the best fit values
of $A$ obtained by D18 (see Table 2 of that paper). As explained in 
section 3, assuming that the outflows of each source are powered at 
least in part by the spin of the black hole, it is shown that eq. (2) 
can be applied to FP sources. The same sets of equations used to obtain 
eq. (2) may be manipulated to solve for the accretion disk magnetic field 
strength, resulting in eqs. (7) and (8). The accretion disk magnetic field 
strengths are independent of the beam power. 
A more intuitive derivation of the key equations is provided in section 3.2; here the 
empirical nature of the method is explicitly demonstrated. 

In the context of the outflow method, black hole spin indicators and indicators
of the accretion disk properties are obtained without
specifying an accretion disk emission model, and without specifying a model relating
beam power to compact radio source luminosity. 
Thus, the accretion disk magnetic field strengths obtained may be used to study and 
constrain accretion disk emission models, and the empirically determined relationship 
between beam power and compact radio luminosity may be used to study and 
constrain jet launching and emission models, as discussed in section
3.3. In addition, the dimensionless field strength provides an
estimate of the mass accretion rate, as discussed in sections 3.1 and 3.3. 

The GBH sample of S15 and the AGN samples of M03, NB16, and D16 are used 
to obtain spin functions, spin values, and magnetic field strengths for each 
measurement of the beam power, bolometric disk luminosity, and black hole mass.
This requires that values of the normalization factors $g_{bol}$ and 
$g_j$ be specified. Fig. 1 of D18 suggests values of $g_{bol} = 1$, 
and $g_j = 0.1$, and these values are adopted here. Tables 2 - 5 list
all quantities of interest, and spin indicators and accretion 
disk magnetic field strengths are summarized in Table 1.  

Reliable and independent spin determinations are available for one GBH and 
6 AGN, and there is good agreement between values obtained here and published 
values; the comparisons are listed in Table 1. The remaining spin values
obtained here could be considered predictions, and indicate relatively
high values for M87 and Sgr $A^*$ (see Table 1); values for all sources 
are included in Tables (2-5). There are several factors that could affect these 
values, such as those discussed in sections 1.1, 4, and 5.  

The spin values obtained are similar for all types of AGN studied and 
for the GBH studied, suggesting that spin value does not determine 
AGN type for the types of sources studied here. The distributions of accretion disk
magnetic field strengths in physical units for the AGN are broad and similar for
different AGN types, and peak at a value
of about $10^4$ G; those for GBH have distributions that peak at a
value of about $10^8$ G. 

Multiple measurements of a particular source provide a unique
opportunity to study and test the outflow method of determining source
parameters. The GBH studied here allow the use of simultaneous or 
contemporaneous radio and X-ray data to be used to 
study the accretion disk magnetic field strength and black hole spin of
one source at different times. The empirical determinations indicate 
that the variation of the disk field strength modifies the beam power 
but does not affect the empirically determined black hole spin. 
The fact that contemporaneous radio and X-ray data yield a fixed
spin but a variable field strength provides support for the method 
and suggests that it provides an accurate description of the types of 
sources studied here. 

The spin values obtained here are relatively high, 
with typical spin values ranging
from about 0.6 to 1. This could be a selection effect, since all of the 
sources studied have powerful collimated outflows, and sources with 
low spin may produce weak outflows. There is a hint of this in the current
analysis. Considering the AGN samples, the spin values of the M03
sources extend to lower values 
than those for NB16 and D16, and the M03 sample includes many lower
luminosity radio sources. In addition, in the context of the models considered 
here, an outflow requires both an accretion event to provide the magnetized 
plasma to thread the ergosphere of the black hole, and a spinning black hole 
to provide an energy source to power the outflow.

\section*{Acknowledgments}

It is a pleasure to thank the many colleagues with whom this work was
discussed, especially Rychard Bouwens, Jean Brodie,  Ray Carlberg, 
Martin Haehnelt, Zoltan Haiman, Garth Illingworth, Massimo Ricotti, 
Michele Trenti, and Rosie Wyse. Thanks are extended to Chris Reynolds 
and to the referee for providing very helpful comments and 
suggestions. This work was was supported in part by Penn 
State University and performed in part at the Aspen 
Center for Physics, which is supported by National Science Foundation
grant PHY-1607611.  
 
\vfill 
\eject

\newpage

\begin{longtable}[h!]{l|l|l|l|l|l|l|l} 

\caption{Beam Power, Spin Parameters, and Disk Magnetic Field Strength Estimates for 102 Measurements of 4 GBH; all luminosities are in erg/s}   
  \label{tab:table2}\\
  \toprule
(1)&(2)&(3)&(4)&(5)&(6)&(7)&(8)\\
&$\rm{Log}$&$\rm{Log}$&$\rm{Log}$&$\rm{Log}$&&$\rm{Log}$&$\rm{Log}$\\
Source 	&$(L_{Edd})$&$(L_j)$&$(L_{bol})$&$(\sqrt{{f(j) \over f_{max}}})$&
$j	$&$(B/B_{Edd})	$&$(B/10^4 G)	$\\
\hline
  \endfirsthead 
  \toprule
(1)&(2)&(3)&(4)&(5)&(6)&(7)&(8)\\
&$\rm{Log}$&$\rm{Log}$&$\rm{Log}$&$\rm{Log}$&&$\rm{Log}$&$\rm{Log}$\\
Source 	&$(L_{Edd})$&$(L_j)$&$(L_{bol})$&$(\sqrt{{f(j) \over f_{max}}})$&
$j	$&$(B/B_{Edd})	$&$(B/10^4 G)	$\\
\hline
\endhead 						
GX	$	339-4	$&$	38.89	$&$	37.08	$&$	38.06	$&$	-0.21	$&$	0.89	$&$	-0.19	$&$	4.19	$\\
GX	$	339-4	$&$	38.89	$&$	37.08	$&$	38.07	$&$	-0.21	$&$	0.89	$&$	-0.19	$&$	4.19	$\\
GX	$	339-4	$&$	38.89	$&$	37.02	$&$	37.64	$&$	-0.14	$&$	0.95	$&$	-0.29	$&$	4.09	$\\
GX	$	339-4	$&$	38.89	$&$	37.02	$&$	37.51	$&$	-0.11	$&$	0.97	$&$	-0.33	$&$	4.06	$\\
GX	$	339-4	$&$	38.89	$&$	36.62	$&$	36.94	$&$	-0.18	$&$	0.92	$&$	-0.46	$&$	3.93	$\\
GX	$	339-4	$&$	38.89	$&$	36.73	$&$	37.44	$&$	-0.24	$&$	0.87	$&$	-0.34	$&$	4.05	$\\
GX	$	339-4	$&$	38.89	$&$	37.05	$&$	38.02	$&$	-0.22	$&$	0.89	$&$	-0.20	$&$	4.18	$\\
GX	$	339-4	$&$	38.89	$&$	37.06	$&$	38.04	$&$	-0.22	$&$	0.89	$&$	-0.20	$&$	4.19	$\\
\hline 
\end{longtable}

\begin{longtable}[h!]{l|l|l|l|l|l|l|l|l|l} 
\caption{Beam Power, Spin Parameters, and Disk Magnetic Field Strength Estimates for 80 Compact Radio Sources; all luminosities are in erg/s. }  % title of Table
  \label{tab:table3}\\
  \toprule
(1)&(2)&(3)&(4)&(5)&(6)&(7)&(8)&(9)&(10)\\
&&D&$\rm{Log}$&$\rm{Log}$&$\rm{Log}$&$\rm{Log}$&&$\rm{Log}$&$\rm{Log}$\\
Source	& 	Type	&(Mpc)&$(L_{Edd})$&$(L_j)$&$(L_{bol})$&
$\sqrt{({f(j) \over f_{max}})}$&$j$&$(B/B_{Edd})$&$(B/10^4 G)$\\
\hline
  \endfirsthead
  \toprule
(1)&(2)&(3)&(4)&(5)&(6)&(7)&(8)&(9)&(10)\\
&&D&$\rm{Log}$&$\rm{Log}$&$\rm{Log}$&$\rm{Log}$&&$\rm{Log}$&$\rm{Log}$\\
Source	& 	Type	&(Mpc)&$(L_{Edd})$&$(L_j)$&$(L_{bol})$&
$\sqrt{({f(j) \over f_{max}})}$&$j$&$(B/B_{Edd})$&$(B/10^4 G)$\\

\hline
\endhead 
Ark	$	564	$&	NS1	&$	106	$&$	44.14	$&$	43.61	$&$	44.98	$&$	0.06	$&$	1.00	$&$	0.17	$&$	1.93	$\\

Cyg	$	A	$&	S2/L2	&$	240	$&$	47.51	$&$	45.67	$&$	45.46	$&$	0.00	$&$	1.00	$&$	-0.42	$&$	-0.35	$\\

IC	$	1459	$&	L2	&$	31	$&$	47.11	$&$	44.42	$&$	41.90	$&$	0.22	$&$	1.00	$&$	-1.07	$&$	-0.80	$\\

IC	$	4296	$&	L1.9	&$	201	$&$	47.21	$&$	44.40	$&$	42.44	$&$	0.07	$&$	1.00	$&$	-0.98	$&$	-0.76	$\\

IC	$	4329A	$&	S1	&$	70	$&$	44.80	$&$	43.86	$&$	44.96	$&$	0.00	$&$	1.00	$&$	0.03	$&$	1.46	$\\

Mrk	$	3	$&	S2	&$	56	$&$	46.92	$&$	44.53	$&$	43.94	$&$	-0.09	$&$	0.98	$&$	-0.61	$&$	-0.24	$\\

Mrk	$	279	$&	S1.5	&$	136	$&$	45.73	$&$	43.75	$&$	44.96	$&$	-0.34	$&$	0.76	$&$	-0.16	$&$	0.80	$\\

Mrk	$	335	$&	NS1	&$	115	$&$	45.25	$&$	43.38	$&$	44.54	$&$	-0.29	$&$	0.81	$&$	-0.15	$&$	1.06	$\\
\hline 
\end{longtable}

\begin{longtable}[h!]{l|l|l|l|l|l|l|l|l|l} 
\caption{Beam Power, Spin Parameters, and Disk Magnetic Field Strength 
Estimates for 100 FRII AGN including the 97 D16 sources plus 3 W sources 
from Grimes et al. (2004); all luminosities are in erg/s.}    % title of Table
  \label{tab:table4}\\
  \toprule
(1)&(2)&(3)&(4)&(5)&(6)&(7)&(8)&(9)&(10)\\
&&&$\rm{Log}$&$\rm{Log}$&$\rm{Log}$&$\rm{Log}$&&$\rm{Log}$&$\rm{Log}$\\
Source 	$	$&Type	&$z 	$&$(L_{Edd})$&$(L_j)$&$(L_{bol})$&
$(\sqrt{{f(j) \over f_{max}}})$&$j$&$(B/B_{Edd})	$&$(B/10^4 G)$\\
\hline
  \endfirsthead % <-- This denotes the end of the header, which will be shown on the first page only
  \toprule
(1)&(2)&(3)&(4)&(5)&(6)&(7)&(8)&(9)&(10)\\
&&&$\rm{Log}$&$\rm{Log}$&$\rm{Log}$&$\rm{Log}$&&$\rm{Log}$&$\rm{Log}$\\
Source 	$	$&Type	&$z 	$&$(L_{Edd})$&$(L_j)$&$(L_{bol})$&
$(\sqrt{{f(j) \over f_{max}}})$&$j$&$(B/B_{Edd})	$&$(B/10^4 G)$\\
\hline
\endhead % <-- Everything between \endfirsthead and \endhead will be shown as a header on every page
%\scriptsize
3C	$	33	$&	HEG	&$	0.059	$&$	46.68	$&$	44.39	$&$	45.63	$&$	-0.42	$&$	0.67	$&$	-0.23	$&$	0.27	$\\

3C	$	192	$&	HEG	&$	0.059	$&$	46.52	$&$	44.05	$&$	45.22	$&$	-0.45	$&$	0.63	$&$	-0.29	$&$	0.29	$\\

3C	$	285	$&	HEG	&$	0.079	$&$	46.64	$&$	44.04	$&$	44.36	$&$	-0.30	$&$	0.80	$&$	-0.50	$&$	0.01	$\\

3C	$	452	$&	HEG	&$	0.081	$&$	46.83	$&$	44.62	$&$	45.08	$&$	-0.22	$&$	0.88	$&$	-0.38	$&$	0.04	$\\

3C	$	388	$&	HEG	&$	0.09	$&$	46.95	$&$	44.42	$&$	44.13	$&$	-0.15	$&$	0.94	$&$	-0.62	$&$	-0.26	$\\

3C	$	321	$&	HEG	&$	0.096	$&$	46.90	$&$	44.24	$&$	45.79	$&$	-0.59	$&$	0.48	$&$	-0.24	$&$	0.14	$\\

3C	$	433	$&	HEG	&$	0.101	$&$	47.08	$&$	44.80	$&$	44.38	$&$	-0.05	$&$	0.99	$&$	-0.59	$&$	-0.30	$\\

3C	$	20	$&	HEG	&$	0.174	$&$	46.69	$&$	45.11	$&$	44.75	$&$	0.14	$&$	1.00	$&$	-0.43	$&$	0.06	$\\
\hline 
\end{longtable}

\begin{longtable}[h!]{l|l|l|l|l|l|l|l|l} 
\caption{Beam Power, Spin Parameters, and Disk Magnetic Field Strength Estimates for 576 LINERS; all luminosities are in erg/s}
   % title of Table
  \label{tab:table5}\\
  \toprule
(1)&(2)&(3)&(4)&(5)&(6)&(7)&(8)&(9)\\
Source&&$\rm{Log}$&$\rm{Log}$&$\rm{Log}$&$\rm{Log}$&&$\rm{Log}$&$\rm{Log}$\\
NB	&$z	$&$(L_{Edd})	$&$(L_j)$&$(L_{bol})	$&
$(\sqrt{{f(j) \over f_{max}}})$ &$j$&$(B/B_{Edd})$&$(B/10^4 G)	$\\
\hline
  \endfirsthead 
  \toprule
(1)&(2)&(3)&(4)&(5)&(6)&(7)&(8)&(9)\\
Source&&$\rm{Log}$&$\rm{Log}$&$\rm{Log}$&$\rm{Log}$&&$\rm{Log}$&$\rm{Log}$\\
NB	&$z	$&$(L_{Edd})	$&$(L_j)$&$(L_{bol})	$&
$(\sqrt{{f(j) \over f_{max}}})$ &$j$&$(B/B_{Edd})$&$(B/10^4 G)	$\\

\hline
\endhead 
1	&$	0.0608	$&$	45.67	$&$	43.16	$&$	42.97	$&$	-0.17	$&$	0.93	$&$	-0.58	$&$	0.42	$\\

2	&$	0.1117	$&$	46.39	$&$	44.55	$&$	44.80	$&$	-0.08	$&$	0.98	$&$	-0.34	$&$	0.30	$\\

3	&$	0.1628	$&$	45.95	$&$	43.66	$&$	43.51	$&$	-0.12	$&$	0.96	$&$	-0.53	$&$	0.33	$\\

4	&$	0.1316	$&$	46.63	$&$	43.51	$&$	43.09	$&$	-0.30	$&$	0.81	$&$	-0.76	$&$	-0.24	$\\

5	&$	0.1319	$&$	45.56	$&$	43.51	$&$	42.47	$&$	0.14	$&$	0.95	$&$	-0.67	$&$	0.39	$\\

6	&$	0.0555	$&$	46.33	$&$	42.93	$&$	42.07	$&$	-0.29	$&$	0.82	$&$	-0.92	$&$	-0.25	$\\

7	&$	0.0536	$&$	45.88	$&$	42.90	$&$	41.19	$&$	0.02	$&$	1.00	$&$	-1.01	$&$	-0.11	$\\

8	&$	0.0503	$&$	45.54	$&$	42.86	$&$	42.18	$&$	-0.12	$&$	0.96	$&$	-0.72	$&$	0.34	$\\
\hline 
\end{longtable}
\end{document}